\documentclass[structabstract]{aa}
\usepackage{graphicx}
\usepackage{txfonts}
\usepackage{natbib}
\bibpunct{(}{)}{;}{a}{}{,}

\usepackage{longtable}
\usepackage{amssymb}
\usepackage[latin1]{inputenc}

\def\kms{km~s$^{-1}\,$}
\def\G23{G23.01--0.41}
\def\Jyb{Jy~beam$^{-1}$}

\begin{document}
   \title{VLBI study of maser kinematics in high-mass SFRs. \textrm{II}. G23.01--0.41}


  \author{ A. Sanna \inst{1,} \inst{2}
\and L. Moscadelli \inst{3} \and R. Cesaroni \inst{3} \and A. Tarchi \inst{2} \and R. S. Furuya \inst{4} \and C. Goddi \inst{5,} \inst{6}}
   \offprints{A. Sanna, \email{asanna@ca.astro.it}}

   \institute{Dipartimento di Fisica, Università degli Studi di Cagliari, S.P. Monserrato-Sestu km 0.7, I-09042 Cagliari, Italy
   \and INAF, Osservatorio Astronomico di Cagliari, Loc. Poggio dei Pini, Strada 54, 09012 Capoterra (CA), Italy
   \and INAF, Osservatorio Astrofisico di Arcetri, Largo E. Fermi 5, 50125 Firenze, Italy
   \and Subaru Telescope, National Astronomical Observatory of Japan, 650 North A'ohoku Place, Hilo, HI 96720, USA
   \and European Southern Observatory, Karl-Schwarzschild-Strasse 2, D-85748 Garching bei M$\ddot{\rm u}$nchen, Germany
   \and Harvard-Smithsonian Center for Astrophysics, 60 Garden Street, Cambridge, MA 02138, USA}



  \abstract
   {}
   {We performed a detailed study of maser and radio continuum emission toward the high-mass star-forming region \G23. This study aims at improving our knowledge of the high-mass star-forming process by comparing the gas kinematics near a newly born young stellar object (YSO), analyzed through  high spatial resolution maser data, with the large-scale environment of its native hot molecular core (HMC), identified in previous interferometric observations of thermal continuum and molecular lines.}
   {\textbf{Using the VLBA and the EVN arrays, we conducted phase-referenced observations of the three most powerful maser species in \G23: H$_2$O at 22.2~GHz (4 epochs), CH$_3$OH at 6.7~GHz (3 epochs), and OH at 1.665~GHz (1 epoch). In addition, we performed high-resolution ($\geq0\farcs1$), high-sensitivity ($<0.1$~mJy) VLA observations  of the radio continuum emission from the HMC at 1.3 and 3.6~cm.}}
   {We have detected H$_2$O, CH$_3$OH, and OH maser emission clustered within 2000 AU from the center of a flattened HMC, oriented SE-NW, from which emerges a massive $^{12}$CO outflow, elongated NE-SW, extended up to the pc-scale.
Although the three maser species show a clearly different spatial and velocity distribution and sample distinct environments around the massive YSO, the spatial symmetry and velocity field of each maser specie can be explained in terms of expansion from a common center, which possibly denotes the position of the YSO driving the maser motion.
Water masers trace both a fast shock (up to 50~\kms) closer to the YSO, powered by a wide-angle wind, and a slower (20~\kms) bipolar jet, at the base of the large-scale outflow.
Since the compact free-free emission is found offset from the putative location of the YSO along a direction consistent with that of the maser jet axis, we interpret the radio continuum in terms of a thermal jet.
The velocity field  of methanol masers can be explained in terms of a composition of slow (4~\kms\ in amplitude) motions of radial expansion and rotation about an axis approximately parallel to the maser jet. Finally, the distribution of line of sight velocities of the hydroxyl masers
suggests that they can trace gas less dense ($\rm n_{H_{2}} \le$~10$^6$~cm$^{-3}$) and more distant from the YSO than that traced by the water and methanol masers, which is expanding toward the observer. A few pairs of OH masers, with different circular polarization, are well aligned in position on the sky and we interpret them as Zeeman pairs. From Zeeman splitting, the derived typical values of the magnetic field are of a few mG.
 }
   {}


   \maketitle
%

\section{Introduction}

Hot, dense, molecular cores (HMCs) of dust and gas located in giant molecular clouds (GMCs) are the birth sites of high-mass young stellar objects (YSOs). Ultraviolet light from the newly formed stars ionizes the surrounding gas, creating a compact H~\textsc{ii} region (e.g., \citealt{Hoare2007}), excites several molecular transitions (e.g., \citealt{Sridharan2002}), and heats the dust, which reradiates the energy in the far infrared (FIR) band (e.g., \citealt{Molinari2008}). At the radio wavelengths, maser emissions of several molecular transitions are observed during the earliest evolutionary phases of  high-mass (proto)stars (e.g., \citealt{Szymczak2005}), even before the appearance of an ultra compact H~\textsc{ii} region (UCH~\textsc{ii}). From an observational point of view, however, the study of high-mass star-forming regions (HMSFRs) is challenging because of three main limitations: the dusty environment is obscured at optical/NIR frequencies; time scales of massive star formation are short, and hence, the chance of observing massive YSOs is small; massive YSOs are rare and statistically found far away from the observer, at distances of a few kpc, clustered in tight associations.

Since a few years, we have started an observational campaign to study the high-mass star-forming process by comparing interferometric thermal data, tracing the large-scale environment (e.g., \citealt{Codella1997,Furuya2008}), with Very Long Baseline Interferometry (VLBI) measurements of maser transitions, tracing the inner kinematics of the (proto)stellar cocoon. Details about our VLBI observational program to measure molecular masers in HMSFRs are extensively presented in \citet[][hereafter Paper~\textrm{I}]{Sanna2010}. The present paper focuses on our observations and analysis of the HMSFR \object{G23.01$-$0.41}.

In Sect.~2, we provide an up-to-date review of previous observations toward the HMSFR \G23. Section~3 describes our VLBI observations of the 22.2~GHz water (H$_2$O), 6.7~GHz methanol (CH$_3$OH) and 1.665~GHz hydroxyl (OH) maser transitions, together with the new Very Large Array (VLA) observations of the radio continuum emission at 1.3 and 3.6~cm.
In Sect.~4, we illustrate the spatial morphology, kinematics, and time-variability of individual maser species, and present results from our VLA observations, constraining the properties of the radio continuum observed associated with the masers.
Section~5 discusses the spatial association of the maser species and their overall kinematics, and draws a comprehensive picture of the phenomena observed in the HMSFR \G23 on angular scales from a few mas to tens of arcsec.  Main conclusions are summarized in Sect.~6.

\section{The HMSFR \G23}

The HMSFR \G23 is located at a distance of 4.59$^{+0.38}_{-0.33}$~kpc \citep{Brunthaler2009}, has a large-scale clump mass of about $4 \times 10^3$~M$_\odot$ \citep{Furuya2008}, and a bolometric luminosity of about $10^5$~L$_\odot$ (upper limit) inferred from IR data \citep{Araya2008}\footnote{The reported values of mass and luminosity have been corrected taking into account the accurate distance recently measured by \citet{Brunthaler2009}.}. From observations of NH$_3$ and CH$_3$CN transitions, the systemic velocity of the region (V$_{sys}$) with respect to the local standard of rest (LSR) was determined to be \ 77.4~\kms\ (\citealt{Codella1997,Furuya2008}).

On an arcminute scale, Spitzer IRAC GLIMPSE observations show strong 4.5~$\mu$m excess indicative of shocked gas (\citealt{Araya2008} and references therein), at the position of a molecular clump extended over an area of about \ 30$\arcsec \times 30\arcsec$, imaged with the Nobeyama Millimeter Array (NMA) in the $^{13}$CO~$(1-0)$ and C$^{18}$O~$(1-0)$ lines \citep{Furuya2008}. Associated with the extended mid-IR emission,  a massive $^{12}$CO~$(1-0)$ bipolar outflow was observed, with a prominent red-wing emission, elongated in the NE--SW direction. The same outflow was also detected in the $^{13}$CO~$(1-0)$ and HNCO~$(5_{05}-4_{04})$ lines \citep{Furuya2008}. Single-dish observations toward \G23 detected several, both thermal and maser, molecular transitions.
\citet{Anglada1996} reported similar spatial and velocity distribution of the CS~$(1-0)$ and NH$_3$~$(1,1)$ lines tracing the turbulent, high-density ($10^4-10^5$~cm$^{-3}$) gas in the region (see also \citealt{Larionov1999}). Thermal, broad line (FWHM$ > 20$~\kms) SiO emission in the ground vibrational state (J~=~$2-1$ and~$3-2$) was detected by \citet{Harju1998}.
\citet{Caswell2000} detected thermal CH$_3$OH emission from the 156.6~GHz (see also \citealt{Slysh1999}) and 107.0~GHz transitions, this latter superimposed on a maser component.
Methanol maser emission was also reported at 6.7~GHz (e.g., \citealt{Menten1991,Goedhart2004}), 12.2~GHz (e.g., \citealt{Caswell1993,Blaszkiewicz2004}), 44.1~GHz \citep{Slysh1994}, and 95.2~GHz \citep{Val'tts2000}.
First attempts to measure Zeeman splitting of the 6.7~GHz maser components suggest a magnetic field strength of a few tens of mG \citep{Vlemmings2008}.
The \G23  region hosts both strong 22.2~GHz H$_2$O (e.g., \citealt{Caswell1983b,Szymczak2005}) and 1.665, 1.667 (main-lines), and 1.612~GHz (satellite-line) OH masers (e.g., \citealt{Caswell1983a,Szymczak2004}). Through Zeeman splitting of the 1.667~GHz maser components, \citet{Szymczak2004} estimated a magnetic field strength of a few tenths of mG. Interestingly, in the spectra of the 1.665, 1.667, and 1.720~GHz OH transitions, absorption features were also noted \citep{Caswell1983a, Szymczak2004}. A similar absorption pattern was observed in the 4.8~GHz H$_2$CO transition by \citet{Downes1980}, who associated \G23  to a nearby diffuse H~\textsc{ii} region, identified through the H$110\alpha$ recombination line emission.

On angular scales of a few arcsec, NH$_3$~$(3,3)$ VLA and CH$_3$CN~$(6-5)$ NMA observations \citep{Codella1997, Furuya2008} revealed the presence of an HMC, with an elongated shape and a velocity gradient oriented in the NW--SE direction.
From the 3~mm  dust continuum emission, the HMC mass was estimated to be of about 70~M$_\odot$ \citep{Furuya2008}. These observations suggest that the HMC may be a flattened structure rotating about the axis of the $^{12}$CO~$(1-0)$ bipolar outflow detected at larger scales. The VLA positions of the H$_2$O, OH \citep{FC1999}, and 4.8~GHz H$_2$CO \citep{Araya2008} masers, as well as the positions of
the 6.7~GHz (J.L. Caswell unpublished; \citealt{Caswell2000}) and 12.2~GHz \citep{Brunthaler2009} CH$_3$OH masers  measured with the Australia Telescope Compact Array (ATCA) and Very Long Baseline Array (VLBA), respectively, agree well with the position of the HMC.
From the large collection of multi-wavelength data reported here, the \G23  region can be convincingly depicted as an active site of massive star formation.

\section{Observations and Calibration}

\subsection{VLA observations: 1.3~cm \& 3.6~cm continuum }

The source was observed with the VLA\footnote{The VLA is operated by the National Radio Astronomy Observatory (NRAO). The NRAO is a facility of the National Science Foundation operated under cooperative agreement by Associated Universities, Inc.} at X and K bands in both the C-array (in
April 2008) and A-array configurations (in October 2008). At 3.6~cm in both
configurations and at 1.3~cm in the VLA--C, the continuum mode of
the correlator was used, resulting in an effective bandwidth of 172~MHz. At
1.3~cm in the VLA--A we used mode ``4'' of the correlator, with a
pair of 3.125~MHz bandwidths (64 channels) centered on the strongest H$_2$O
maser line and a pair of 25~MHz bandwidths (8 channels) sufficiently offset
from the maser lines to obtain a measurement of the continuum emission.
The two bandwidths centered at the same frequency (measuring the two circular polarizations)
were averaged.

At X band, 3C\,286 (5.2~Jy) and 3C\,48 (3.1~Jy) were used as primary flux
calibrators, while 1832--105 (1.4~Jy) was the phase calibrator.  For the
K-band observations, the primary flux calibrator was 3C\,286 (2.5~Jy), the
phase calibrator 1832--105 (1.0~Jy), and the bandpass calibrator (for the
VLA--A data only) 1733-130 (3.7~Jy).

The data were calibrated with the NRAO AIPS software package using standard procedures. Only for the
VLA--A data at 1.3~cm, several cycles of self-calibration were applied to
the strongest maser channel, and the resulting phase and amplitude
corrections were eventually transferred to all the other line channels and to
the K-band continuum data. This procedure resulted in a significant (at least
a factor 2) improvement of the signal-to-noise ratio (SNR).

Natural-weighted maps were made with task IMAGR of AIPS both for the
continuum and the line data. Simultaneous observation of the line and
continuum emission at K band made it possible to obtain the relative position
of the continuum image with respect to the H$_2$O maser spots with great
precision ($\sim0\farcs01$). Finally, we identified the maser spots that appear
in both our VLBA and VLA images and, using the accurate absolute position
information obtained from the VLBA observations, we re-centered the VLA spots
and continuum image. The correction thus applied turns out to be $<20$~mas in each coordinate.
In conclusion, we believe that the absolute astrometrical precision of the
continuum maps presented in this paper is of $\sim0\farcs01$.

\subsection{Maser VLBI}

We conducted VLBI observations of the H$_2$O and CH$_3$OH masers (at several epochs), and of the OH maser (at a single epoch) toward \G23 in the K, C, and L bands, respectively. In order to derive the maser absolute position, we used phase-referencing by fast switching between the maser target and the calibrator J1825$-$0737. This calibrator has an angular offset from the maser source of $2\fdg6$ and belongs to the list of sources defining the International Celestial Reference Frame (ICRF). Its absolute position is known to better than $\pm2$~mas and its flux measured with the VLBA at S and X bands is 68 and 151~m\Jyb, respectively \citep{Fomalont2003}.
Five fringe finders (J1642+3948; J1751+0939; J1800+3848; J2101+0341; J2253+1608) were observed for bandpass, single-band delay, and instrumental phase-offset calibration.
Data were reduced with AIPS following the VLBI spectral line procedures.
Paper~\textrm{I} describes the general data calibration procedures.

\subsubsection{VLBA observations: 22.2~GHz H$_2$O masers}

\begin{figure*}
\centering
\includegraphics [width=7cm]{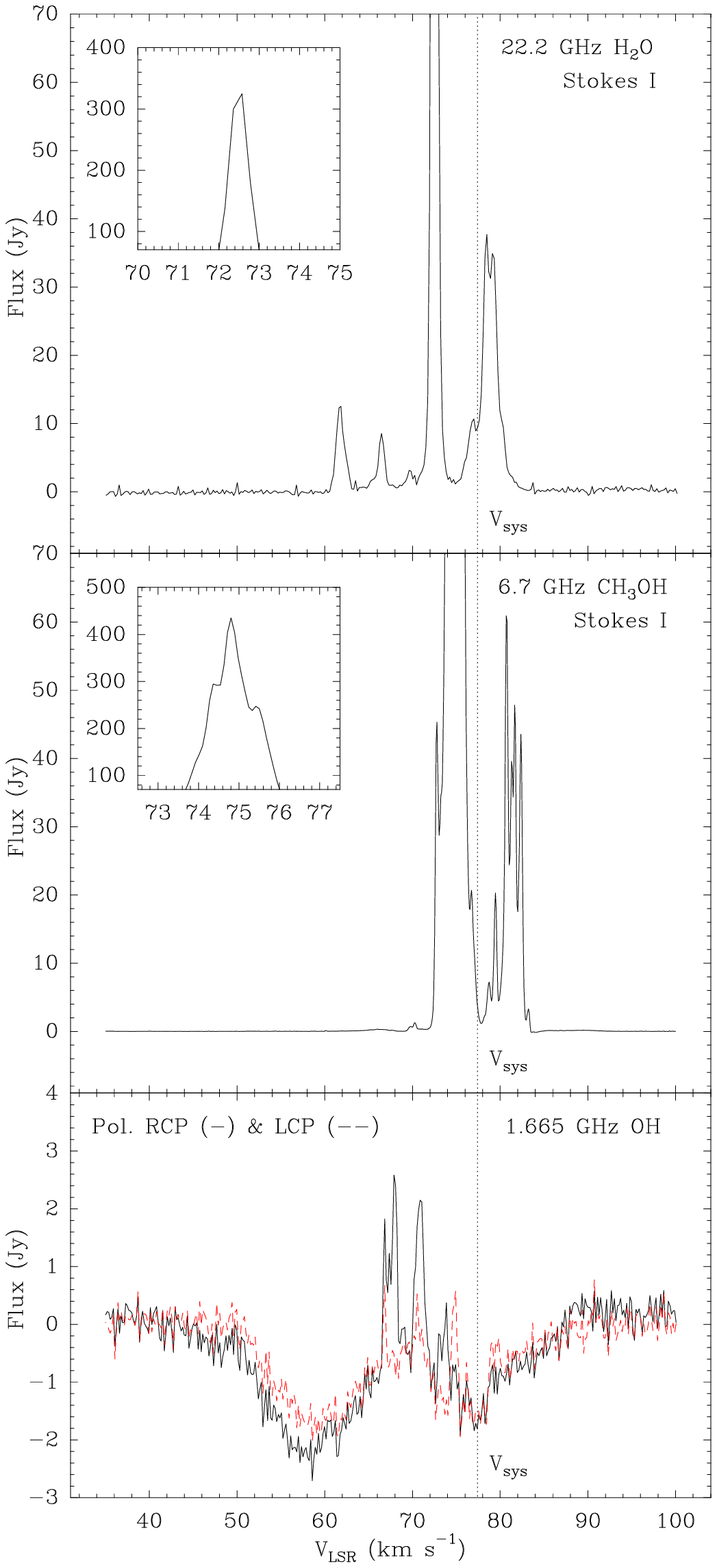}
\caption{Total-power spectra of the H$_2$O, CH$_3$OH and OH masers toward \G23. \emph{Upper panel:} system-temperature (T$_{sys}$) weighted average of the 22.2~GHz total-power spectra of the 9 VLBA antennas observing on 2007 January 17 (BR, FD, KP, LA, MK, NL, OV, PT, SC). Inset: zoom of the water maser peak. \emph{Middle panel:} Effelsberg total-power spectrum of the 6.7~GHz methanol maser emission on 2007 March 17. Inset: zoom of the methanol maser peak. \emph{Lower panel:} T$_{sys}$-weighted average of the 1.665~GHz total-power spectra of the 9 VLBA antennas observing on 2007 April 27 (BR, FD, HN, KP, MK, NL, OV, PT, SC). Continuous (black) and dashed (red) lines are used to distinguish between right (RCP) and left (LCP) circular polarizations, respectively. The dotted line crossing the spectra represents the systemic velocity (V$_{sys}$) inferred from CH$_3$CN measurements.}
\label{fig1}
\end{figure*}

We observed the HMSFR \G23 (tracking center: $R.A.$(J2000)~$=18^h34^m40\fs39$ and $Dec.$(J2000)~$=-09\degr00'38\farcs5$) with the VLBA\footnote{The VLBA is operated by the NRAO.} in the $6_{16}-5_{23}$ H$_2$O transition (rest frequency 22.235079~GHz). The observations (program code: BM244) consisted of 4 epochs: April 17, June 29, and September 22, 2006, and January 17, 2007. During a run of about 6~h per epoch, we recorded the dual circular polarization through a 16~MHz bandwidth centered on a LSR velocity (V$_{\rm LSR}$) of  77.0~\kms. The data were processed with the VLBA FX correlator in Socorro (New Mexico) using an averaging time of 1~s and 1024 spectral channels. The total-power spectrum of the 22.2~GHz masers toward \G23 is shown in Fig.~\ref{fig1} (top panel). This profile was obtained by averaging the
total-power spectra of all VLBA antennas, weighting each spectrum with the antenna system temperature (T$_{sys}$).

The natural CLEAN beam was an elliptical Gaussian with a FWHM size of about $ 1.3~\textrm{mas} \times 0.4~\textrm{mas}$ at a P.A. of $-19\degr$ (east of north), with little variations from epoch to epoch. The interferometer instantaneous field of view was limited to about 2\farcs7. At each observing epoch, using an on-source integration time of about 2.5~h, the effective rms noise level of the channel maps ($\sigma$) varied in the range \ 0.006--0.03~\Jyb. The spectral resolution was 0.2~\kms.

\subsubsection{VLBA observations: 1.665~GHz OH masers}

We observed the HMSFR \G23 (tracking center: $R.A.$(J2000)~$=18^h34^m40\fs27$ and $Dec.$(J2000)~$=-09\degr00'37\farcs5$) with the VLBA in the $^2\Pi_{3/2}$ $\rm J=3/2$ OH transition (rest frequency 1.665401~GHz), on  April 27, 2007 (program code: BM244N). During a run of about 6~h, we recorded the dual circular polarization through two bandwidths of 1~MHz and 4~MHz, both centered at a LSR velocity of 70.0~\kms.
The 4~MHz bandwidth was used to increase the SNR of the weak L-band
signal of the continuum calibrator. The data were processed with the VLBA FX correlator in two correlation passes using 1024 and 512 spectral channels for the 1~MHz and 4~MHz bands, respectively. In each correlator pass, the data averaging time was 2~s. The T$_{sys}$-weighted mean of antenna total-power spectra for the right and left circular polarizations at 1.665~GHz are shown in Fig.~\ref{fig1} (bottom panel). After removing a first order baseline, the 1.665~GHz OH spectra of both circular polarizations present similar characteristics, i.e. several maser emission components superposed on a broad absorption feature (see Sect.~\ref{OH_results}).

The natural CLEAN beam was an elliptical Gaussian with a FWHM size of $19~\textrm{mas} \times 10~\textrm{mas}$ at a P.A. of $3\degr$. The interferometer instantaneous field of view was limited to about 18\farcs5. With an on-source integration time of about 1.9~h, the effective rms noise level of the channel maps was about 0.02~\Jyb. The 1~MHz band spectral resolution was 0.2~\kms. The visibility phase of both circular polarizations was calibrated using as phase-reference the brightest maser channel of the right circular polarization.

\subsubsection{EVN observations: 6.7~GHz CH$_3$OH masers}

We observed the HMSFR \G23 (tracking center: $R.A.$(J2000)~$=18^h34^m40\fs39$ and $Dec.$(J2000)~$=-09\degr00'38\farcs5$) with the European VLBI Network (EVN)\footnote{The European VLBI Network is a joint facility of European, Chinese and other radio astronomy institutes founded by their national research councils.} in the $5_{1}-6_{0}A^+$ CH$_3$OH transition (rest frequency 6.668519~GHz).
This work is based on 3 epochs (program codes: EM061, EM069), separated by about 1~yr, observed
on February 27, 2006, on March 17, 2007, and on March 16, 2008. At the first two epochs, antennas involved in the observations were Cambridge, Jodrell2, Effelsberg, Hartebeesthoek, Medicina, Noto, Torun, and Westerbork.
Since the longest baselines involving the Hartebeesthoek antenna (e.g., Ef-Hh baseline about 8042~km) heavily resolve the maser emission and do not produce fringe-fit solutions, the Hartebeesthoek antenna was replaced with the Onsala antenna in the third epoch.
During a run of about 6~h per epoch, we recorded the dual circular polarization through two bandwidths of 2~MHz and 16~MHz, both centered on a LSR velocity of 77.0~\kms. The 16~MHz bandwidth was useful to increase the SNR of the weak continuum calibrator. The data were processed with the MKIV correlator at the Joint Institute for VLBI in Europe (JIVE - Dwingeloo, The Netherlands) using an averaging time of 1~s and 1024 spectral channels for both observing bandwidths. The Effelsberg total-power spectrum at 6.7~GHz toward \G23 is shown in Fig.~\ref{fig1} (middle panel).

The natural CLEAN beam was an elliptical Gaussian with a FWHM size of about $ 13~\textrm{mas} \times 5~\textrm{mas}$ at a P.A. of $30\degr$, slightly varying from epoch to epoch. The interferometer instantaneous field of view was limited to about 9\farcs2. At each observing epoch, using an on-source integration time of about 2.2~h, the effective rms noise level of the channel maps varied in the range \ 0.008--0.3~\Jyb. The 2~MHz band spectral resolution was 0.09~\kms.

\section{Results}

The present section reports our main results for the radio continuum and individual maser species in \G23. We describe in Paper~\textrm{I} the criteria used to identify, derive parameters (position, intensity, flux and size), and measure proper motions of individual masing-clouds. In the following, the term ``spot'' refers to maser emission on a single velocity channel, whereas the term ``feature'' indicates a collection of
spots emitting at similar position over contiguous channels (i.e. an individual masing-cloud).

\subsection{1.3~cm \& 3.6~cm continuum }

\begin{figure*}
\centering
\includegraphics [angle=-90,width=7cm]{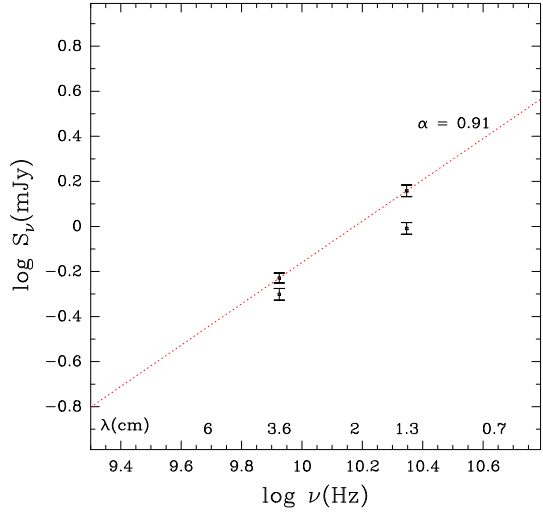}
\caption{Spectral energy distribution of the radio continuum in the HMSFR \G23. Dots and errorbars report the values and the associated errors (1$\sigma$) of our measurements at 1.3 and 3.6~cm using both the VLA--A and VLA--C configurations (see Table~\ref{tab4}). Taking at each observing frequency the highest-flux measurement, the derived spectral index, indicated by the dotted line, is \ $\alpha$ = 0.9 .}
\label{fig6}
\end{figure*}

Our measurements of the 1.3 and 3.6~cm continuum emission in both VLA--A  and VLA--C configurations are summarized in Table~\ref{tab4}.
While the most extended VLA configuration partly resolved the radio continuum at 1.3~cm (missing about half of the flux measured in the compact configuration), the fluxes at 3.6~cm of both configurations are comparable. Taking into account the shape of the VLA--A beam (HPBW of $0\farcs13 \times 0\farcs09$ at a P.A. of $-13\degr$), the 1.3~cm continuum emission appears slightly elongated in the E--W direction (see Fig.~\ref{fig2}). Fig.~\ref{fig6} shows that a spectral index ($\alpha$) of 0.9 is derived from our highest-flux measurements.

Previous observations at 3 and 6~cm were performed using the ATCA and the VLA--BnA configuration, respectively \citep{F&C2000,Araya2008}. \citet{F&C2000} reported a flux at 3~cm of 0.37$\pm 0.12$~mJy with an HPBW of $1\farcs5 \times 1\farcs5$, and \citet{Araya2008} derived a $5\sigma$ upper limit of 2.4~mJy in the 6~cm band (HPBW of $1\farcs6 \times 1\farcs1$ at a P.A. of $-7\degr$).

\subsection{22.2~GHz H$_2$O masers}\label{H2O_results}

We imaged the whole range of LSR velocities, from 68 to 87~\kms, where \citet{FC1999} detected water maser emission. Using the VLA--C they identified two regions of maser activities, each elongated in the north-south direction by about $1''$, and separated by about $40''$. Our field of view covers the region associated with their brightest spots and the absolute position we have determined for the reference maser spot is offset by about 1\farcs5 from the reference position of \citet{FC1999}.

We have identified 86 distinct water maser features distributed within an area of about $ 1\arcsec \times 0\farcs5 $. They are mainly grouped in
three clusters to the NE, SW, and towards the center of the plotted area, which are identified in Fig.~\ref{fig2}a with labels ``A'', ``B'', and ``C'', respectively. The individual features properties are presented in Table~\ref{wat_tab}. Their intensities range from 0.07 to 202.89~\Jyb. Only 28 features (33\% of the total) persisted over at least 3 epochs, 15 of which lasting 4 epochs. The spread in LSR velocities ranges from 89.4~\kms, for the most redshifted feature~(\#~57), to 57.0~\kms, for the most blueshifted one (\#~22). The clusters ``A'' and ``B'' are observed at a similar angular separation of about 0\farcs5 (i.e. $2 \times 10^3$~AU) from the cluster ``C'' and both have a threadlike N--S morphology extending up to 117~mas. Water maser emission from cluster ``A'' (15 features) and ``B'' (6 features)  presents different properties: the cluster ``B'' consists of faint (weaker than 0.49~\Jyb), short-living (one epoch) features, whereas the cluster ``A'' consists of bright (up to 22.21~\Jyb) and persistent emission (67\% of the features lasted at least 3~epochs and 8 features persisted 4 epochs). The cluster ``C'' has an arc-like morphology, of about 200~mas in size, and extends on top of the 1.3~cm continuum emission. Water maser emission from this region (61 features) is the most variable: only 25\% of the observed features persisted for at least 3 epochs and only 4 features lasted 4 epochs. This region has also undergone a powerful maser flare (feature \#~1 in Table~\ref{wat_tab}) observed during the last epoch (Fig.~\ref{fig1} upper panel) with an increase in intensity of about 200 times.
The water masers absolute distribution and line of sight (l.o.s.) velocities are plotted in Fig.~\ref{fig2}a. Using a time baseline of 9~months, we measured relative transverse velocities of individual features with a mean accuracy of about 30\%. The magnitude of relative proper motions ranges from
6.0$\pm$3.7~\kms, for feature~\#~21, to 59.3$\pm$4.7~\kms, for feature~\#~6. Positions and velocities are relative to the isolated and compact feature~\#~24. Our measurements of absolute proper motions have been corrected by the velocity vector $\mu_x = -37$ and $\mu_y = -90$~\kms (where the $X$ and $Y$ indices refer to the motion components toward the east and the north, respectively) representing the Galactic peculiar motion of the HMSFR as measured by \citet{Brunthaler2009}. Mean uncertainties in the magnitude of the absolute motions of water maser features are about 40\%. The relative and absolute proper motions of water masers are plotted in Figs.~\ref{fig4}a~and~\ref{fig4}b, respectively.
Absolute velocities of maser features belonging to cluster ``C'' show a regular variation of orientation with position, rotating from north to south
through the west across the arc-like distribution of features, and it appears that they are diverging from a point to the east of the maser cluster.

\subsection{1.665~GHz OH masers \& absorption}\label{OH_results}

We imaged the whole range of LSR velocities, from 65 to 81~\kms, where 1.665~GHz OH maser emission was detected by \citet{FC1999}. Using the VLA A-B hybrid configuration they found OH maser emission scattered over a region of about $ 2\arcsec \times 1\arcsec $. The absolute position of our reference maser spot is offset by about 0\farcs9 from the reference position of \citet{FC1999}.

Although the whole distribution of hydroxyl masers is scattered across a region of about $ 0\farcs8 \times 0\farcs4 $, half of maser features are clustered into a small area, $0\farcs06 \times 0\farcs12$, eastward of the 1.3~cm continuum peak (Fig.~\ref{fig2}c). We have identified 24 distinct OH maser features and individual features properties are presented in Table~\ref{hyd_tab}. Positions are relative to the isolated and bright feature~\#~1. The absolute distribution and l.o.s. velocities of hydroxyl masers are plotted in Fig.~\ref{fig2}c.  The LSR velocities range from 79.7~\kms, for the most redshifted feature~(\#~13), to 60.5~\kms, for the most blueshifted one (\#~23). The mean FWHM line width of individual maser features is 0.3~\kms. The l.o.s. velocities are blueshifted toward the radio continuum and closer to V$_{sys}$ at larger distances. An ordered velocity shift from 73.0 to 62.6~\kms\ (features with label number 8, 11, 15, 12, 17, 20, 16, 22) is observed in an elongated structure extending N--S for 33~mas ($\approx 150$~AU) over the central region. Maser intensities range from 0.12 to 2.39~\Jyb\, with the brightest features showing blueshifted l.o.s. velocities. We have identified 5 Zeeman pairs satisfying the condition that the circular polarization components coincide positionally within 3$\sigma$ and are separated in velocity by more than their typical line width of 0.3~\kms, corresponding to magnetic fields less than 0.5~mG (e.g., inset in Fig.~\ref{fig2}c; see also \citealt{Fish2006}). The l.o.s. magnetic field strength ranges from -5.8 to +3.6~mG. OH maser total-power spectra show an absorption feature superposed to the maser emission (Fig.~\ref{fig1} bottom panel). The absorption feature extends over about 40~\kms, from about 50 to 90~\kms, with two deeps at the LSR velocities of 58.5 and 77.3~\kms. Absorption is not visible on the cross-power spectra of the shortest VLBA baseline (KP--PT $\approx 400$~km). Assuming the absorption is resolved out, this allows to set a lower limit of about 0\farcs09 to the size of the absorption region.

\subsection{6.7~GHz CH$_3$OH masers}\label{CH3OH_results}

We imaged the whole range of LSR velocities, from 69 to 84~\kms, where 6.7~GHz maser emission was previously detected towards
\G23 by \citet{Caswell1995} using the Parkes 64-m telescope. The ATCA position of the 6.7~GHz maser peak reported by \citet{Caswell2000} falls inside the area of our 6.7~GHz maser detections.

\begin{figure*}[htbp]
\centering
\includegraphics[angle=0.0,scale=1.0]{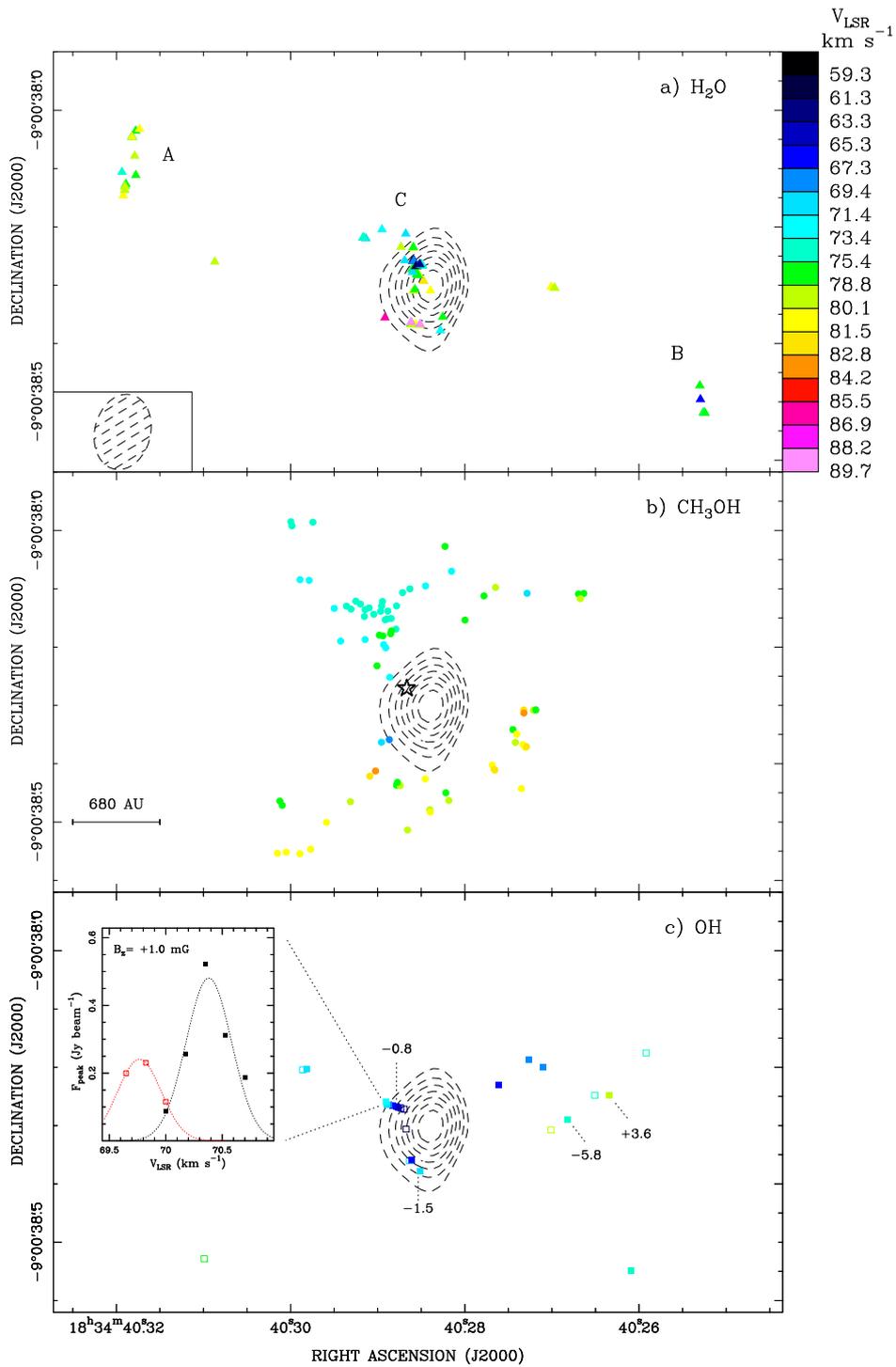}
\caption{Absolute positions and LSR velocities of the three maser species observed in \G23:  \emph{a)} H$_2$O (triangles), \emph{b)} CH$_3$OH (dots), and \emph{c)} OH (squares). Different colors are used to indicate the maser LSR velocities, according to the color scale on the right-hand side of the plot, with green representing the systemic velocity of the HMC. The VLA 1.3~cm continuum emission is plotted with dashed contours. Contour levels range from 30 to $90\%$ of the peak emission (0.72~m\Jyb) at multiples of $10\%$. The restoring beam is shown in the lower left corner of the upper panel. The linear scale of the plots is shown in the lower left corner of the middle panel. For water masers, letters A, B, and C indicate the clusters defined in Sect.~\ref{H2O_results}. For methanol masers, the star marks the position of the barycenter of the methanol maser distribution, as defined in Sect.~\ref{CH3OH_results}.  For hydroxyl masers, full and empty squares denote right and left circularly polarized features, respectively, and numbers close to (some) maser features indicate the inferred local value of the magnetic field strength (in mG). \emph{Inset:} example of Zeeman pair for the feature \#~11, reporting the Gaussian fits to the spectral profiles of the RCP (black line) and LCP (red line) components. The RCP and LCP measurements are indicated with full and empty squares, respectively.}
\label{fig2}
\end{figure*}

Methanol maser emission at 6.7~GHz is distributed within a region of about $ 0\farcs6 \times 0\farcs6 $ about the 1.3~cm continuum emission (Fig.~\ref{fig2}b). We have identified 81 distinct maser features that are grouped depending on the l.o.s. velocity: redshifted features lie to the south of the radio continuum and blueshifted ones to the north. The individual feature properties are presented in Table~\ref{met_tab}. Maser LSR velocities range from 83.2~\kms, for the most redshifted feature~(\#~38), to 67.8~\kms, for the most blueshifted one (\#~81). The average offsets in l.o.s. velocities (with respect to V$_{sys}$) of the maser features belonging to the southern and northern regions of the radio continuum are  \ $+4.4$ and \ $-2.1$~km~s$^{-1}$, respectively.
Maser intensities are on average very bright (30\% of the features have peak emission greater than 10~\Jyb), but range from 0.10 to 146.03~\Jyb.
67 features (83\% of the total) persisted over the 2~yr interval covered by our 3 observing epochs.
For each maser feature, Table~\ref{met_tab} reports the mean brightness variability defined as the ratio between the variation of the brightness (of the strongest spot) and the average of the minimum and maximum brightness.
The mean brightness variability of methanol maser features was about 45\% (compared to only 20\% for the source G16.59--0.05, Paper~\textrm{I}), varying from a minimum of 5\%, for feature~\#~80, to a maximum of about 136\%, for feature~\#~48. These estimates of brightness variability should be taken as upper limits since they include effects from the slightly varying beam shape among different epochs as well as amplitude calibration uncertainties. Positions are relative to the structurally-stable, compact, and bright feature~\#~10, whose intensity varies less than about 6\%. Methanol maser positions and l.o.s. velocities are plotted in Fig.~\ref{fig2}b.
We have calculated the geometric center of \emph{all the detected} features (hereafter ``barycenter'', identified with label~S in Table~\ref{met_tab}), whose position is indicated by a star in Fig.~\ref{fig2}b. In deriving the ``barycenter'' position, to reduce the weight of clustered maser features, uniform weighting has been used, weighting feature positions with the reciprocal of the local feature density. Note that, since \emph{only a subset} of 6.7~GHz maser features had a stable (spatial and spectral) structure over time, the ``barycenter'' is not a good point to refer the internal motions.
We have also calculated the geometric center (hereafter ``center of motion'', identified with label~\#~0 in Table~\ref{met_tab}) of features with a stable spatial and spectral structure persisting over the 3 observing epochs, and refer our measurement of transverse motions to this point (Fig.~\ref{fig5}a).
Using a 2~yr time baseline, the relative transverse velocities of individual maser features are determined with a mean accuracy of about 30\%.
Proper motions, derived only for maser features with a stable structure, are calculated by performing a linear fit of feature position vs. time. To establish the extent of accelerated motions, an analysis of deviations from linear motion is postponed after an upcoming fourth epoch. The magnitude of relative proper motions ranges from 1.7$\pm$1~\kms, for feature~\#~17, to 16.0$\pm$1.6~\kms, for feature~\#~15. Transverse velocities of maser features have a mean value of 7.2~\kms, about 2--3 times the spread of l.o.s. velocities. We observe a cluster of maser features connected by a bridge of extended weaker emission ($110~\textrm{mas} \times 60~\textrm{mas}$), emitting over a narrow velocity range of a few~\kms\ (Fig.~\ref{fig5}b). Since in this region spatial blending of nearby maser centers limits the precision of positions derived with Gaussian fitting,
proper motions from this area were sufficiently accurate only for the brightest, more isolated features (\#~1, 4, 9, 17, 18, 19).

Looking at Fig.~\ref{fig2}b, 6.7~GHz maser features seem to trace a bipolar, funnel-like structure, centered to the east of the 1.3~cm continuum peak. Maser proper motions presented in Fig.~\ref{fig5}a describe a smoothly varying velocity field, roughly symmetrical with respect to the barycenter of the spatial distribution. The regularity and near symmetry of the maser velocity distribution makes us confident that the calculated proper motions are representative of the true motions of the methanol masing-clouds.

\section{Discussion}\label{discussion}

\subsection{Nature of the radio continuum} \label{nat_rad}

The continuum source appears slightly resolved  at the VLA--A 1.3~cm angular resolution of $\sim0\farcs1$ (see Figs.~\ref{fig3}--\ref{fig5}).
The fluxes of the continuum source at 1.3~and~3.6~cm (see Table~\ref{tab4}) are consistent with free-free emission from
an UCH~\textsc{ii} region with radius of about 450~AU ionized by a Lyman continuum of \ 1--3 $\times$ 10$^{45}$~s$^{-1}$,
corresponding to a ZAMS B1 star \citep{Schraml1969,Panagia1973}. The spectral index \ $\alpha$ = 0.9 between 3.6~and~1.3~cm could be interpreted with the emission becoming optically thin at \ $\lambda \la 1$~cm.

An alternative interpretation for the weak continuum emission is in terms of a shock induced ionization in a thermal jet. Following the
calculation by \citet{Ang96}, for optically thin emission, one has: $ \rm F \, d^2 = 10^{3.5} \, (\Omega/4\pi) \, \dot{P} $, where \ F is the measured
continuum flux in mJy, $\dot{\rm P}$ is the outflow momentum rate in \mbox{M$_{\sun}$~yr$^{-1}$~\kms}, $\Omega$ is the jet solid angle in sr,
and \ d is the source distance in kpc.  Using the flux of 1.4~mJy measured at 1.3~cm with the VLA--C, and the accurate distance of 4.6~kpc to
\G23 by \citet{Brunthaler2009}, one derives:   $ \dot{\rm P} = 10^{-2} \, (\Omega/4\pi)^{-1}$~\mbox{M$_{\sun}$~yr$^{-1}$~\kms}.
The momentum rate thus depends on the estimate of the jet collimation factor  \ $(\Omega/4\pi)^{-1}$. If the jet structure is not resolved but only a bright knot along the jet axis is observed, the jet solid angle can be estimated from the angular size of the knot if the position of the YSO powering the jet is known. In the next section
we discuss the maser information with the aim to constrain the YSO position.

\subsection{Water masers tracing a wind/jet system}\label{wat_dis}

Current models explain H$_2$O maser excitation by collisional pumping with H$_2$ molecules within hot ($\gtrsim 400$~K) shocked layers of gas behind both high-velocity ($\geq$~50~km~s$^{-1}$) dissociative (J-type; \citealt{Elitzur1989}) and slow ($\leq$~45~km~s$^{-1}$) non-dissociative (C-type; \citealt{Kaufman1996}) shocks, propagating in dense regions (H$_{2}$ pre-shock density in the range \ $ 10^7$--$10^9$~cm$^{-3}$). Following these models, the cluster ``C'' of water masers witnesses the presence of dense shocked gas near the continuum source. The higher variability of the bright water masers from this cluster, compared with clusters ``A'' and ``B'', could indicate that the cluster ``C'' pinpoints the most active site of the region. The arc-like distribution of maser features in this region, together with their fast (as high as 55~\kms) and diverging proper motions (see Fig.~\ref{fig4}b), lead us to think that this water maser emission traces a shock front expanding from a center located to the east of the 1.3~cm continuum peak. The barycenter of the funnel-like structure of the 6.7~GHz masers (the star in Fig.~\ref{fig2}) falls also in this area, and we speculate that is the place of the massive YSO(s), possibly a multiple system, exciting and driving both maser emissions. In the following discussion, we will use the 6.7~GHz maser barycenter as the best guess for the YSO(s) position.

If water maser emission is excited in a jet from a YSO, knowing the average distance of water masers from the
YSO and the average maser velocity, one can estimate the momentum rate of the jet assuming that is momentum driven. Using the model-predicted H$_{2}$ pre-shock density of \ $10^8$~cm$^{-3}$, the momentum rate in the water maser jet is given by the expression:
$ \rm \dot{P} = 1.5 \times 10^{-3} \, V_{10}^{2} \, R_{100}^{2} \, (\Omega/4\pi)$~\mbox{M$_{\sun}$~yr$^{-1}$~\kms}, where \ $\rm V_{10}$ \ is the average maser velocity in units of 10~\kms, \ $\rm R_{100}$ \ is the average distance of water masers from the YSO in units of 100~AU, and \ $\Omega$ is the solid angle of the jet. This expression has been calculated by multiplying the momentum rate per unit surface transferred to the ambient gas (n$_{H_2}$ m$_{H_2}$ V$^2$), by a factor \ $\Omega \rm R^2$, assuming that the jet is emitted from a source at a distance R from the masers within a beaming angle \ $\Omega$.

We hypothesize that a  wind emitted from the YSO impacts against the dense circumstellar gas, and excites and drives the motion of water masers of cluster ``C''.  Water masers of cluster ``C'' have an average (sky-projected) distance from the YSO of $\approx200$~AU, and their average absolute
velocity is 40~\kms. Assuming an isotropic wind ($\Omega = 4\pi $), and considering that the derived sky-projected distance should be taken as a lower limit to the
true distance, we derive a value for the momentum rate \
$ \dot{\rm P} \ge 0.1$~\mbox{M$_{\sun}$~yr$^{-1}$~\kms}. This value is one order of magnitude higher than the typical momentum
rate of the stellar wind from a ZAMS O star of  \ 10$^{-2}$~\mbox{M$_{\sun}$~yr$^{-1}$~\kms} \citep{Mar05}, and might be explained if the massive YSO
was still accreting mass and a fraction of mass approaching the YSO was deflected into the wind, as predicted by X-wind \citep{Shu00}
and disk-wind \citep{Kon00} models.

Assuming that the position of the YSO is close to the geometric center of the (methanol and water) maser distributions, the 1.3~cm continuum source is found offset from the YSO by about 60~mas to the SW. Such an offset is significantly larger than the accuracy ($\approx10$~mas) of the
absolute position of the continuum source, which leads us to think that the best interpretation is in terms of a
thermal jet rather than an UCH~\textsc{ii} region. We describe a case similar to what observed for the well-studied source IRAS~20126+4104, where
the YSO position is clearly offset of the peaks of the VLA 3.6~cm continuum emission, tracing shock induced ionization in a thermal jet \citep{Hof07}.
Towards \G23, the VLA--A slightly resolves the 1.3~cm emission, which indicates that the source size is less than the VLA--A beam at 1.3~cm of $\approx110$~mas.
Then, at the distance of the 1.3~cm continuum peak from the YSO
of $\approx60$~mas, the jet has to be collimated within a cone of semi-aperture \ $\theta \le 0.5 \times 110/60 = 0.9$~rad,
corresponding to a jet solid angle $\Omega \le 2.5$~sr. Using this value in the expression for the momentum rate derived in
Sect.~\ref{nat_rad}, one has \  $ \dot{\rm P} \ge 5 \times 10^{-2}$~\mbox{M$_{\sun}$~yr$^{-1}$~\kms}, which is consistent with the
value of the momentum rate derived above for the YSO wind.

Figure~\ref{fig2}a shows that water masers \emph{not belonging} to cluster ``C'' are distributed to the NE (mainly in cluster ``A'') and to the SW
(mainly in cluster ``B'') of the putative location of the YSO.
They could trace shocked gas along the path of the bipolar jet, emerging from the YSO and interacting with the
surrounding molecular core. Such a jet could be the engine of the massive $^{12}$CO~$(1-0)$ bipolar outflow, elongated in the NE--SW direction,
observed on arcsec scales (see the sketch in Fig.~\ref{fig8}). This interpretation explains the absolute velocities of water masers in cluster ``A'' oriented towards NE (see Fig.~\ref{fig4}b). The same jet should be responsible for the excitation of the continuum source and all the water maser features separated from the YSO by more than $\approx0\farcs2$. To intercept all the water masers found at larger distance from the YSO,
the jet semi-aperture has to be larger than \ $\theta \ge 0.25$~rad, which is consistent with the condition \ $\theta \le 0.9$~rad
from the upper limit to the angular size of the continuum source.

Water masers in cluster ``A'' move with an average velocity of 20~\kms\ and are separated from the YSO by $\approx2300$~AU.
Using these values in the expression for the momentum rate in the maser jet, one finds: $ \dot{\rm P} = 3.2 \, (\Omega/4\pi) $~\mbox{M$_{\sun}$~yr$^{-1}$~\kms}. If the same jet is responsible for exciting the continuum source and driving the water masers
of cluster ``A'', the jet solid angle ($\Omega$) can be determined by requiring that the momentum rate in the maser jet equals that from the continuum emission:
$$ 3.2 \, (\Omega/4\pi)  = 10^{-2} \, (\Omega/4\pi)^{-1} $$
\noindent From this equation one derives \  $ \Omega = 0.7$~sr \ corresponding to a semi-opening angle of the jet \ $\theta = 0.5$~rad.
This value, derived from measurements of the continuum flux and the maser kinematics of cluster ``A'',
is consistent with the independent condition \ $ 0.25 \le \theta \le 0.9$ \ owing to the spatial distribution of all detected water masers and to the size of the continuum source. That supports our result. From the derived value of the jet solid angle, the
jet momentum rate is $  \dot{\rm P} = 0.2$~\mbox{M$_{\sun}$~yr$^{-1}$~\kms}.

This value for the jet momentum rate agrees with that (0.1~\mbox{M$_{\sun}$~yr$^{-1}$~\kms}) of the dominant (SW redshifted) lobe of the large-scale $^{12}$CO~$(1-0)$ outflow reported by \citet{Furuya2008}. These evidence support
the interpretation that the NE--SW distribution of water masers traces the jet at the base of  the massive molecular outflow observed
at arcsec scales. Note also that the momentum rate in the jet is comparable with the momentum rate (0.1~\mbox{M$_{\sun}$~yr$^{-1}$~\kms}) derived for the wide-angle wind driving the water masers of cluster ``C'', which presents the case that the jet itself is powered by the YSO wind. At a distance of $\approx300$~AU from the YSO, i.e. at the position of the 1.3~cm source, most of the momentum of the wind would be efficiently collimated within the jet, with a collimation factor \ $(\Omega/4\pi)^{-1} \approx 20$. For comparison, values of collimation factors of the order of 10 are typical for well-studied radio jets in low-~and~intermediate-mass star-forming regions  \citep{Ang96}. If the high momentum rate of the wind could hint to the YSO being still accreting mass, the efficient collimation of the wind into a jet could indicate that the accretion onto the YSO is mediated through an accretion disk.

\subsection{Methanol maser environment}\label{met_dis}

\begin{figure*}[htbp]
\centering
\includegraphics[angle=0.0,scale=1.1]{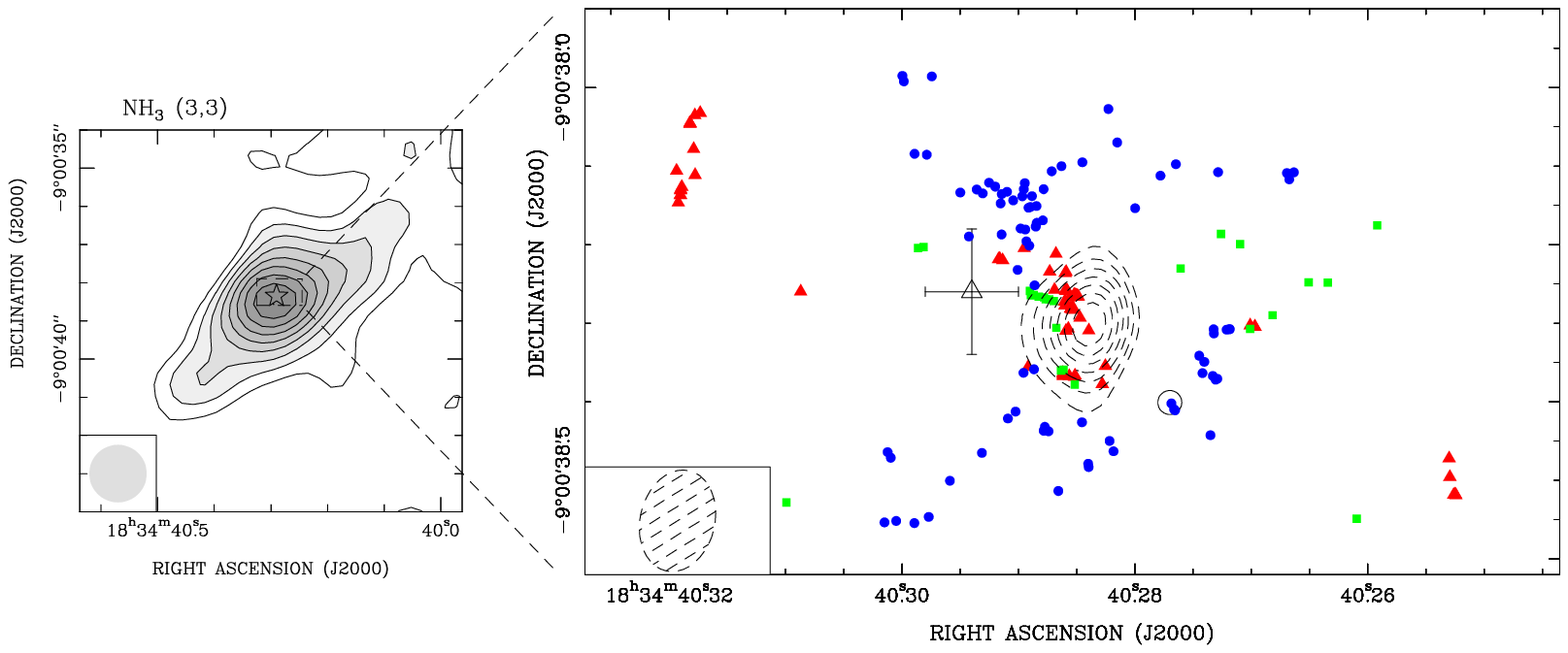}
\caption{Collection of the sub-arcsec observations toward \G23. \emph{Left panel:} map in the NH$_3$ (3,3) line from \citet{Codella1997}. Contour levels range from 20 to $90\%$ of the peak emission (46~m\Jyb) at multiples of $10\%$. The restoring beam is shown in the lower left corner of the panel. The star marks the peak position of the 3~mm continuum emission as determined by \citet{Furuya2008}. \emph{Right panel:} enlargement of the region over which we have detected maser emission. Red triangles, blue dots, and green squares represent respectively H$_2$O, CH$_3$OH, and OH maser positions from our VLBI measurements. The empty circle marks the position of the brightest 12.2~GHz CH$_3$OH maser feature (at V$_{\rm LSR} = 81.5$~\kms) measured by \citet{Brunthaler2009} with an uncertainty of a few mas. The empty triangle with the errorbars indicates the position (and the associated uncertainty) of the 4.8~GHz H$_2$CO maser feature (at V$_{\rm LSR} = 73.6$~\kms) derived by \citet{Araya2008}. Dashed contours show the 1.3~cm VLA continuum emission. Contour levels range from 30 to $90\%$ of the peak emission (0.72~m\Jyb) at multiples of $10\%$. The restoring beam is shown in the lower left corner of the panel.}
\label{fig3}
\end{figure*}

\begin{figure*}[htbp]
\centering
\includegraphics[angle=0.0,scale=1.3]{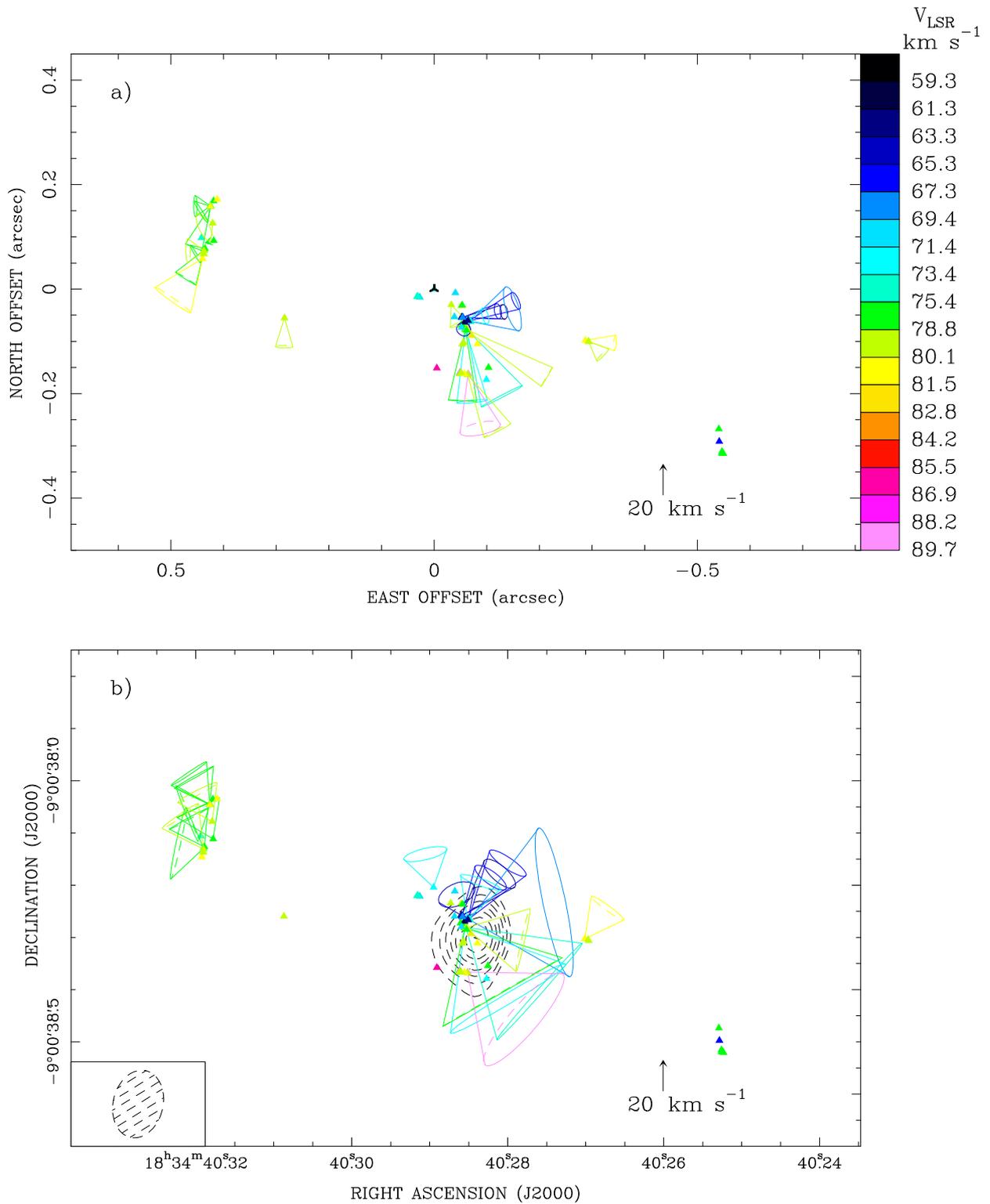}
\caption{22.2~GHz H$_2$O maser  kinematics toward \G23. \emph{a)} positions (triangles) and transverse velocities of the H$_2$O maser features relative to the feature~\#~24 (indicated with the vertex-connected symbol). Colored cones are used to show both the direction and the uncertainty (cone aperture) of the proper motion of maser features. The proper motion amplitude scale is given by the black arrow on the bottom right corner of the panel. Proper motions of the maser features~\#~30, 38, and 62 are too uncertain and, thus, not plotted. Different colors are used to indicate the maser LSR velocities, according to the color scale on the right-hand side of the panel, with green denoting the systemic velocity of the HMC. \emph{b)} absolute positions and transverse velocities of the water maser features. The potted field of view is the same as in the upper panel and symbols have the same meaning as in the upper panel. Dashed contours show the VLA 1.3~cm continuum emission. Contour levels range from 30 to $90\%$ of the peak emission (0.72~m\Jyb) at multiples of $10\%$. The restoring beam is shown in the lower left corner of the panel.}
\label{fig4}
\end{figure*}

\begin{figure*}[htbp]
\centering
\includegraphics[angle=0.0,scale=1.3]{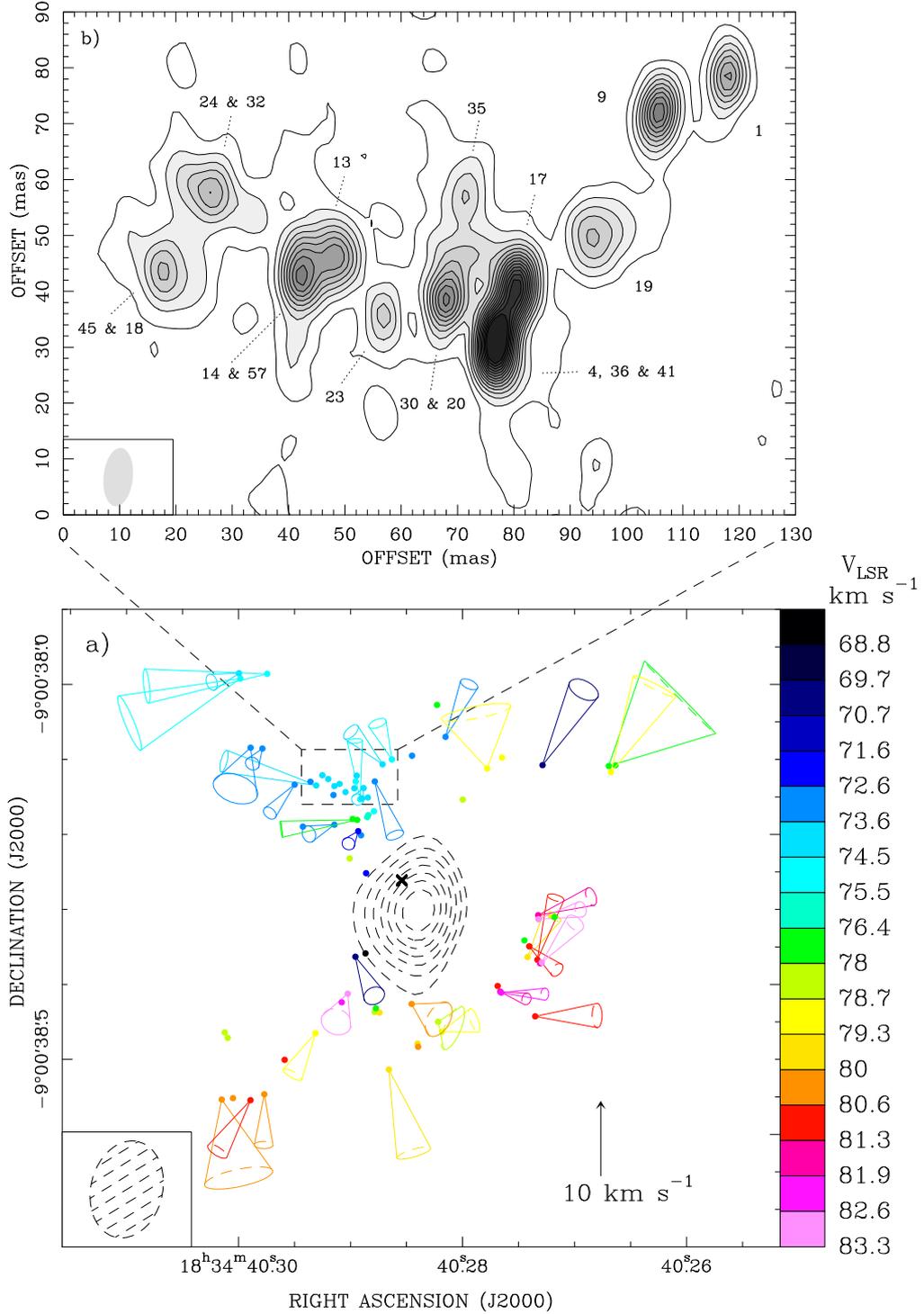}
\caption{6.7~GHz CH$_3$OH maser kinematics toward \G23. \emph{a)} absolute positions (dots) and transverse velocities of the CH$_3$OH maser features relative to the center of motion (as defined in Sect.~\ref{CH3OH_results}) of the methanol maser distribution (indicated by the cross). Colored cones are used to show both the direction and the uncertainty (cone aperture) of the proper motion of maser features. The proper motion amplitude scale is given by the black arrow on the bottom right corner of the panel. Different colors are used to indicate the maser LSR velocities, according to the color scale on the right-hand side of the plot, with green denoting the systemic velocity of the HMC. Dashed contours show the VLA 1.3~cm continuum emission. Contour levels range from 30 to $90\%$ of the peak emission (0.72~m\Jyb) at multiples of $10\%$. The restoring beam is shown in the lower left corner of the panel. \emph{b)} Zoom of the crowdest region of 6.7~GHz maser emission. The map is obtained from data of the second EVN epoch (2007 March 17), summing the maser emission over the velocity range from 73.40 to 74.63~\kms. Plotted contour levels are spaced by factor of 2, with the lowest level corresponding to the 3$\sigma$ rms of 3.0~\Jyb. The restoring beam is given in the lower left corner of the panel. The numbers near emission peaks are the maser feature labels given in Table~\ref{met_tab}.}
\label{fig5}
\end{figure*}

In Sect.~\ref{wat_dis}, we have used the symmetries in the spatial distributions of the 6.7~GHz methanol and 22~GHz water masers to constrain the YSO position, and postulated that the YSO is located close to the geometric center of both maser species distributions.
Figure~\ref{fig3} shows that 6.7~GHz and 22.2~GHz masers emerge from nearby but \emph{different} positions around the putative location of the massive YSO. Figures~\ref{fig4}b and~\ref{fig5}a show that water masers move significantly faster than methanol masers, and the average direction of motion of the two maser species is also different. Therefore, the comparison of the spatial and velocity distribution suggests that the two maser species, although associated with the same massive YSO, originate from distinct environments, characterized by a different kinematics and, likely, as predicted by water and methanol maser excitation models, by different physical conditions (e.g., \citealt{Moscadelli2007, Goddi2007}).
CH$_3$OH excitation models \citep{Cragg2005} predict that Class~II methanol maser emission is produced by radiative pumping in cool ($\approx 30$~K), or moderately warm regions ($< 200$~K), with H$_{2}$ densities in the range \ $10^6$--$10^9$~cm$^{-3}$. Even though comparably high densities are required for both water and Class~II methanol maser action, strong 6.7~GHz methanol masers can be produced at gas temperature significantly lower than that of the shocked layers of gas emitting water masers.

\begin{figure*}
\centering
\includegraphics[angle=0,width=7cm]{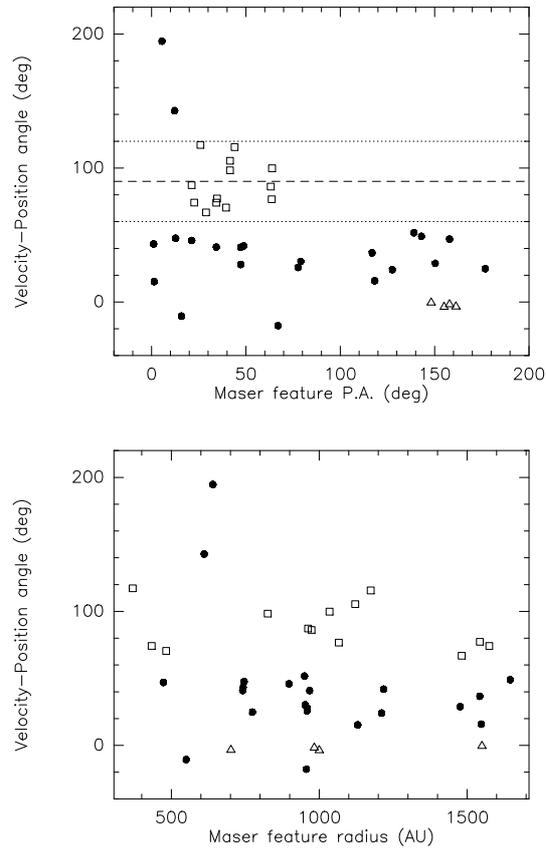}
\caption{Upper and lower panels show the distribution of the angle between the proper motion and the position
vectors of 6.7~GHz masers as a function of the maser P.A. and radial distance, respectively. Approximately azimuthal and
radial velocities are indicated with empty squares and triangles, respectively (velocity-position angle in the range from 60$\degr$ to 120$\degr$  and from -10$\degr$ to 10$\degr$, respectively), whereas filled dots are used for other values of the velocity-position angle.
The 6.7~GHz maser barycenter (see Table~\ref{met_tab}) is taken to be the origin of the (sky-projected) maser positions. The angle between the velocity and the position vectors of a given maser feature is evaluated positive counterclockwise and is equal to the difference between the P.A.
of the feature velocity and position vectors. Maser features with P.A. in the range \ $180\degr -360\degr$
have been folded into the P.A. interval \ $0\degr -180\degr$ (by subtracting 180$\degr$ to their true P.A.). In the upper plot, horizontal dotted and dashed lines mark the boundary and the center, respectively, of the interval over which maser velocities are approximately azimuthal.}
\label{fig7}
\end{figure*}

Looking at Fig.~\ref{fig5}a, one notes that the distribution of 6.7~GHz maser velocities is not isotropic about the maser barycenter.
Taking the maser barycenter as the origin of a system of polar coordinates, (sky-projected) maser velocities with a dominant radial component
appear to concentrate at a P.A. different from that where maser velocities are predominantly azimuthal. Figure~\ref{fig7} shows the distribution of the angle between the maser proper motion and the corresponding position vectors, as a function of both the radial distance and the P.A. of the maser feature. Most of maser velocities form an angle with the radius (outward from the maser barycenter) varying in the range from -20$\degr$ to 120$\degr$ (evaluated positive counterclockwise for the observer). While the distribution of the velocity-position angle appears to be rather uniform with the maser radial distance, all the azimuthal velocities (forming with the radius an angle within \ $90\degr\pm30\degr$) concentrate in a well defined P.A. range, $20\degr \le$ P.A. $\le 70\degr$.

\begin{figure*}
\centering
\includegraphics [angle=0,scale=0.5]{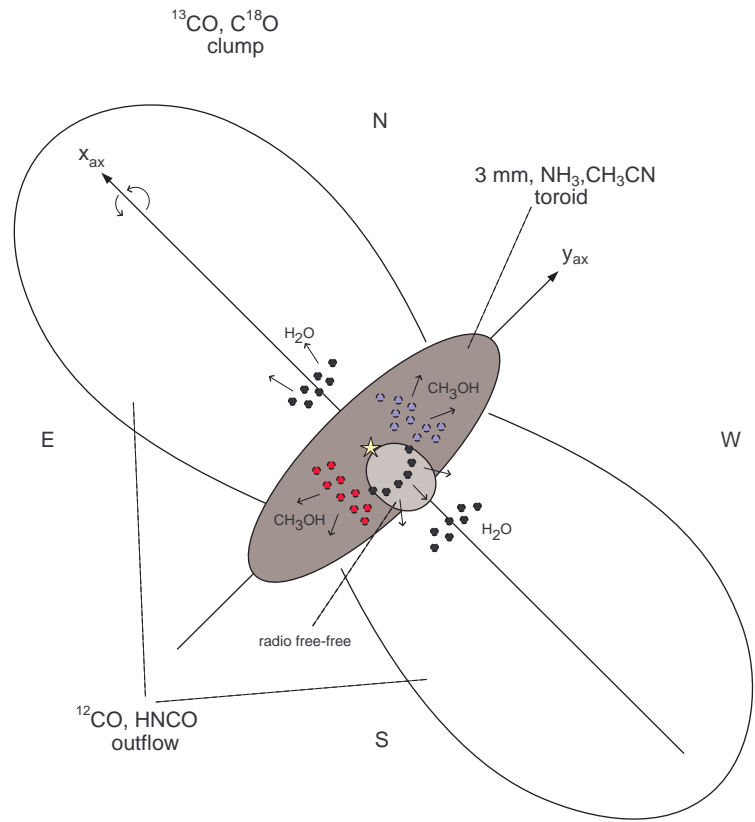}
\caption{Schematic cartoon of the main components in the molecular cloud associated with \G23. The drawing is not to scale. The components represented are (i) the outflow/jet structure traced by CO and H$_2$O maser emission (black dots) from pc-scale to a few thousand of AU; (ii) the toroid structure seen on a large  scale in CH$_3$CN and highly excited NH$_3$ molecules, and its inner portion traced by CH$_3$OH maser emission (blue and red dots according to the l.o.s. velocity); (iii) the radio continuum emission superposed on the arc-like H$_2$O maser structure. The star marks the assumed YSO(s) location.}
\label{fig8}
\end{figure*}

The observed distribution of 6.7~GHz maser velocities can be interpreted in terms of a combination of expanding and rotating motions. Let us consider a rotating structure about an axis inclined with respect to the l.o.s., and indicate with {\bf x$_{\rm ax}$} and {\bf y$_{\rm ax}$} the projections on the plane of the sky of the rotation axis and the line perpendicular to that, respectively (see the sketch in Fig.~\ref{fig8}). Points of the rotating structure belonging to the plane containing the rotation axis and the l.o.s. lie on {\bf x$_{\rm ax}$} and have rotation velocities parallel to {\bf y$_{\rm ax}$}. Therefore, it is possible to identify the projection onto the plane of the sky of the rotation axis looking for the direction in the sky to which the proper motions are perpendicular. The upper plot of Fig.~\ref{fig7} suggests that the azimuthal 6.7~GHz maser velocities can result from rotation around an axis whose {\bf x$_{\rm ax}$} crosses the maser barycenter and is oriented at P.A. within $45\degr\pm25\degr$. Note that this {\bf x$_{\rm ax}$} crosses also the 1.3~cm continuum peak (at a P.A. of $225\degr$) and agrees with the orientation of the 22~GHz water maser jet discussed in Sect.~\ref{wat_dis}. Then, there are indications that 6.7~GHz masers can emerge from a structure rotating around the jet/outflow axis.

Many of the observed 6.7~GHz maser velocities show a component directed radially outward from the maser barycenter, suggesting that the gas
traced by 6.7~GHz masers is also expanding away from the YSO. Looking at Fig.~\ref{fig5}a and the upper plot of Fig.~\ref{fig7}, one notes
that a group of approximately radial velocities (directed within  $10\degr$ from the local radius) concentrates at P.A. between $150\degr - 160\degr$ and $330\degr - 340\degr$, that is along a direction approximately perpendicular to {\bf x$_{\rm ax}$}. If the rotation axis
is close to the plane of the sky, the rotation velocities of the spots belonging to the plane containing the rotation and {\bf y$_{\rm ax}$} axes should be almost parallel to the l.o.s.. Then, the sky-projected velocities along the {\bf y$_{\rm ax}$} axis are expected to show small azimuthal components
and to be mainly radial if the gas is also flowing away from the YSO, in agreement with what we actually observe.

The same assumptions that the 6.7~GHz masers trace gas flowing away from a center and rotating about an axis at small inclination with the plane of the sky, allows also to easily interpret the observed distribution of l.o.s. velocities.
Material on the near and far side of the maser structure would move towards and
away from the observer, respectively. Along the sky-projected rotation axis (i.e. {\bf x$_{\rm ax}$}), the observed blue- and redshifted maser velocities towards the NE and SW, respectively, can be explained if we admit that maser emission to the NE and SW originates from material on the near (approaching) and far (receding) side of the structure, respectively.  Along the direction perpendicular to the sky-projected rotation axis (i.e. {\bf y$_{\rm ax}$}), the distribution of l.o.s. velocities would reflect the sense of rotation, with blue- and redshifted maser velocities to the NW and SE, respectively, in agreement with the observations (Fig.~\ref{fig8}).

If the rotation axis is slightly inclined with respect to the plane of the sky, the small group of maser features along the {\bf y$_{\rm ax}$} axis with approximately radial velocities (negligible azimuthal components),  should be found close to the sky. Then, their distribution of l.o.s. velocities can be used to constrain the central mass, in the hypothesis of centrifugal equilibrium. Their average distance from the maser barycenter (Fig.~\ref{fig7}, lower plot) is $\approx1000$~AU, and the average of the absolute value of their l.o.s. velocities is $\approx4$~\kms. Using these values, a central mass of $\approx20$~M$_{\odot}$ is derived, corresponding to an early~B -- late~O type ZAMS star.

This discussion suggests that the geometrical properties and the amplitude of the observed 6.7~GHz maser velocities can be explained in terms of rotation and expansion about a massive YSO. The idea that the methanol-masing gas can participate in an overall expanding motion has also been proposed by \citet{Bartkiewicz2009} on the basis of single-epoch VLBI observations of a  sample of 31 6.7~GHz maser sites. These authors have noted that most maser regions coincide with a 4.5~$\mu$m emission excess (as it is also the case for \G23), which is commonly interpreted as a tracer of expanding, shocked, molecular gas. In the \G23 region, the derived orientation of the rotation axis (at P.A. within $45\degr\pm25\degr$) is consistent with the scenario where 6.7~GHz CH$_3$OH masers can be tracing the internal portions (within a radius of about 2000~AU from the YSO) of the rotating core, observed in the NH$_3$ and CH$_3$CN lines on arcsec scale along the SE--NW direction \citep{Codella1997, Furuya2008}.
The pattern of CH$_3$CN l.o.s. velocities observed by \citet[][Fig.~11]{Furuya2008}, mainly blueshifted to the NW and redshifted to the SE,
agrees with the distribution of l.o.s. velocities of 6.7~GHz CH$_3$OH masers at much smaller angular scale.
We note that the 12.2~GHz maser emission measured in the \G23 region \citep{Brunthaler2009} has both position and LSR velocity (V$_{\rm LSR} = 81.5$~\kms) consistent with the 6.7~GHz masers (Fig.~\ref{fig3}). In order to produce bright 6.7 and 12.2~GHz masers ($\rm T_b\gtrsim 10^{10}$~K), the Class~II methanol maser pumping models by \citet{Cragg2005} prescribe a combination of warm dust ($\rm T_d=$~175 and 125~K) and cool gas ($\rm T_k=$~30 and 50~K), for a gas density of \ $\rm n_{H_{2}}=$~10$^6$--10$^7$~cm$^{-3}$.
These values agree well with the parameters of the ammonia core ($\rm T_k= $~58~K and $\rm n_{H_{2}}= 6.9 \times 10^6$~cm$^{-3}$) derived by \citet{Codella1997}, and would marginally support the idea of an association of methanol masers with the NH$_3$ core.

The measured pattern of 6.7~GHz maser proper motions offers a simple explanation that
the CH$_3$CN LSR velocity gradient in the toroid mapped by \citet{Furuya2008} presents a not so-well defined orientation.
Gas close to the equatorial plane would indeed undertake a complex motion resulting from a combination of rotation and expansion. Taking all the 6.7~GHz maser features with measured proper motions, the mean absolute value ($\approx4$~\kms) of the radial velocity components is basically the same as that of the azimuthal velocity components, indicating that the masing gas expands and rotates at similar speed.  Assuming that the expansion of the gas traced by the 6.7~GHz masers is momentum driven, the momentum rate to accelerate the gas can be estimated with the expression \ $\rm \dot{P} = M \, V^{2} / R$, where \ M \ is the total mass of the expanding gas, \ V \ the average expansion velocity and \ R \ the radius of the expanding sphere (working in spherical symmetry, for simplicity). Taking an upper limit to the gas density of  \ $\rm n_{H_{2}}=10^7$~cm$^{-3}$, a velocity $\rm V = 4$~\kms, and an upper limit to the observed (sky-projected) distance of 6.7~GHz masers from the YSO of \ $\rm R = 2000$~AU, a
momentum rate  \ $\rm \dot{P} = 3 \times 10^{-3}$~\mbox{M$_{\sun}$~yr$^{-1}$~\kms} \ is derived. This value is about two orders
of magnitude smaller than the momentum rate of the YSO wind supposed to be driving the expansion of water masers of cluster ``C'', indicating that a negligible fraction of the wind, non-collimated into the jet and blowing across the equatorial plane,
could be responsible for the expansion of the gas nearby the YSO. The YSO wind is only one of the possible causes for the observed 6.7~GHz maser outward motion. The spatial sampling of the measured maser velocities is not sufficiently complete for a quantitative
analysis to effectively constrain the geometry of the masing gas and  determine which portions of it is expanding.

\subsection{A molecular expanding layer in front of the central source?}

Figures~\ref{fig2}c~and~\ref{fig3} show that the spatial distribution of the 1.665~GHz OH masers differs from that of  6.7~GHz CH$_3$OH and
22.2~GHz H$_{2}$O masers. The N--S elongated structure traced by the OH masers over the 1.3~cm continuum is slightly offset to the
east of the water maser cluster ``C'', and is found near to the (putative) YSO location. To explain the strongly blueshifted l.o.s. velocities (from $-4$ to $-14$~\kms) of this cluster of OH masers, a possible interpretation is that they are seen in the foreground of the YSO and the radio continuum emission, and that they are originating in a layer of gas moving towards the observer. The same wind, which we have
conjectured to be driving the expansion of water masers in cluster ``C'', might be responsible for accelerating the gas hosting the OH maser structure. The OH and CH$_3$OH Class~II maser excitation models by \citet[][Fig.~4]{Cragg2002} predict strong 1.665~GHz
maser action and inhibition of the 6.7~GHz masers when  \ $\rm n_{H_{2}} \le$~10$^6$~cm$^{-3}$. This suggests that the 1.665~GHz OH masers towards the radio continuum are tracing lower densities than those required for intense 6.7~GHz CH$_3$OH masers. Thus, the properties
of the elongated structure observed (only) in the 1.665~GHz OH maser emission might be explained if the OH masers trace a gas layer with
\ $\rm n_{H_{2}} \le$~10$^6$~cm$^{-3}$, rapidly expanding away from the YSO towards the observer. From Zeeman splitting, the l.o.s. magnetic
field strength in this OH maser layer is estimated to be about 1~mG (see Table~\ref{hyd_tab}, features \#~11~and~20). This result is in general agreement with a detailed VLBA study of OH maser proper motions and polarization properties in several UCH~\textsc{ii} regions by \citet{Fish2007}. These authors found that OH masers trace the expansion of ionized gas and also established a correspondence between the gas density  and the magnetic field, deriving  a value of  the order of 1~mG for the l.o.s. magnetic field strength if \ $\rm n_{H_{2}} = $~10$^6$~cm$^{-3}$.

We propose that the OH absorption is spatially unrelated to the OH maser emission and may instead be connected with the large scale dynamics of the region. In the NRAO VLA Sky Survey, the \G23 region shows diffuse 1.4~GHz radio continuum emission extended on scales of about 1\arcmin.
Toward \G23, using the Effelsberg radiotelescope \citet{Downes1980} reported both diffuse H$110\alpha$ recombination line (4.874~GHz) emission and absorption in the 4.8~GHz H$_2$CO line at various velocities ranging from 55 to 103~\kms. Furthermore, a lower limit of a few arcminutes to the size of the absorption region is  set by the H$_2$CO VLA--D observations by \citet{Araya2008}, who resolved the absorption.
The 1.6~GHz OH absorption feature visible in our VLBA total-power spectra is mainly blueshifted and that may suggest that the absorbing molecular gas stays in front of the extended radio continuum emission and is expanding towards the observer (e.g., \citealt{Szymczak2004}).

\section{Summary and Conclusions}

Using the VLBI technique, we observed the HMSFR \G23 in the three most powerful maser transitions: 22.2~GHz H$_2$O, 6.7~GHz CH$_3$OH, and 1.665~GHz OH. The source \G23 was also observed with the VLA, detecting faint ($\approx$~mJy) radio continuum emission with 0\farcs1 resolution towards the center of the HMC. Our main conclusions can be summarized as follows:

\begin{enumerate}

\item H$_2$O, CH$_3$OH, and OH maser emissions are distributed within  $\approx2000$~AU from the center of a HMC powering a massive bipolar outflow. The three maser species, although associated with the same YSO, present different, although complementary, spatial distributions and kinematical properties.

\item Water masers trace fast (20--50~\kms) outflows emitted from the YSO. An arc-like structure of water masers superposed on the
radio continuum marks a fast shock propagating through dense gas, and is probably driven by a YSO wind. The 1.3~cm continuum source and  the two clusters of water masers, aligned along the NE--SW direction detected at larger distance from the YSO, are likely tracing the jet driving the massive CO outflow observed at larger scales.

\item Methanol masers present a N--S oriented, funnel-like spatial distribution with red- and blueshifted features located to the south and the north, respectively. Observing 3 different EVN epochs spanning 2~yr, we have measured accurate (relative errors $<30\%$) proper motions of 6.7~GHz methanol features. The maser transverse velocities range from a few to about 15~\kms. The pattern of 6.7~GHz maser proper motions
can be interpreted in term of a composition of expansion and rotation around a YSO of about 20~M$_{\odot)}$, with the rotation axis oriented on the sky at similar P.A. to the axis of the jet/outflow system traced by the water masers. It is then plausible that the  CH$_3$OH masers trace the internal portions
of the toroid, elongated along the SE--NW direction, observed in the CH$_3$CN and NH$_3$ lines on arcsec scale.

\item Hydroxyl masers superposed on the radio continuum are probably seen in the foreground and expand outward from the central source tracing a lower density environment than that harboring the methanol and water masers.

\end{enumerate}

This study demonstrates that multi-epoch VLBI observations of different maser species provide information useful to explore the
complex phenomena occurring at distances of tens up to thousands of AU around massive YSOs.
The best scientific return from VLBI maser observations can be expected when such observations will be compared to data of new generation (sub)millimeter interferometers (such as ALMA). Clearly, thermal line observations of both outflow(s) and core(s) with an angular resolution closer to that of VLBI maser data will help to solve ambiguities in the interpretation of kinematic structures, thanks to a better sampling of the (proto)stellar environment and its physical properties.

\begin{acknowledgements}

This work is partially supported by a Grant-in-Aid from the Ministry of Education,
Culture, Sports, Science and Technology of Japan (No. 20740113).

\end{acknowledgements}

\bibliographystyle{aa}
\bibliography{SANNApaperG23}

\clearpage

\clearpage

\addtocounter{table}{0}
\begin{table*}
\caption{G23.01--0.41: Radio continuum emission. \label{tab4}}

\begin{tabular}{c c c c c c c c c}

\hline \hline
          &            &              &           &                   & \multicolumn{2}{c}{ Peak position$^a$ } &         &            \\
Telescope & $\lambda$  &    HPBW      &    P.A.   &    Image rms      & R.A.(J2000) &  Dec.(J2000) &    F$_{\rm peak}$    &  F$_{\rm int}$ \\
          &    (cm)    & ($'' \times ''$) &  (\degr)  & (mJy beam$^{-1}$) &   (h m s)   &  (\degr $'$ $''$)& (mJy beam$^{-1}$)&   (mJy)    \\

\hline
& & & & & & & & \\
VLA--A     &    1.3     &  $0.131 \times  0.095$  &   -13   &  0.06  &  18 34 40.284  &  -09 00 38.31 &   0.72  & 0.98 \\
VLA--C     &    1.3     &  $1.1   \times  0.8  $  &    0    &  0.05  &    ...         &     ...       &   1.31   & 1.44 \\
VLA--A     &    3.6     &  $0.36  \times  0.24 $  &   -5    &  0.03  &    ...         &     ...       &   0.47  & 0.59 \\
VLA--C     &    3.6     &  $2.7   \times  1.9  $  &    2    &  0.03  &    ...         &     ...       &   0.52  & 0.50 \\

& & & & & & & & \\
& & & & & & & & \\
\hline
& & & & & & & & \\
\end{tabular}

\footnotesize{(a) \
The absolute position of the VLA--A 1.3~cm emission peak, obtained aligning positions of persistent water maser spots between the VLA and VLBA observations, should be accurate to about $\pm~0\farcs01$.\\
}

\end{table*}

\clearpage

\clearpage

\clearpage

\clearpage

\clearpage

\clearpage

\onllongtab{2}{
\begin{longtable}{ccrcrrcrrr}

\caption{\emph{Parameters of VLBA 22.2~GHz water maser features.}
For each identified feature, the label (given in Col.~1) increases with decreasing brightness.
Cols.~2~and~3 report the LSR velocity and brightness of the brightest spot (for the epoch specified by the number in brackets).
Cols.~4~and~5 give the position offset relative to the feature~\#~24, toward the East and the North directions, respectively.
Cols.~6~and~7 report the projected components of the feature proper motion relative to the feature~\#~24, along the East and North directions, respectively. Col.~8 shows the relative error of the proper motion amplitude.
The absolute position and proper motion of the reference feature~\#~24 are reported at the bottom of the Table.} \label{wat_tab} \\

\hline \hline
Feature &  $V_{\rm LSR}$ & \multicolumn{1}{c}{F$_{\rm peak}$}  & & \multicolumn{1}{c}{$\Delta x$} & \multicolumn{1}{c}{$\Delta y$} &  & \multicolumn{1}{c}{$V_{\rm x}$} & \multicolumn{1}{c}{$V_{\rm y}$} & \multicolumn{1}{c}{$\Delta|V|/|V| $} \\
        &  (km s$^{-1}$) & \multicolumn{1}{c}{(Jy beam$^{-1}$)} &  & \multicolumn{1}{c}{(mas)} & \multicolumn{1}{c}{(mas)} &  & \multicolumn{1}{c}{(km s$^{-1}$)} & \multicolumn{1}{c}{(km s$^{-1}$)} &                  \\
\hline
& & & & & & & & &\\
\endfirsthead
\caption{continued.}\\
\hline \hline
Feature &  $V_{\rm LSR}$ & \multicolumn{1}{c}{F$_{\rm peak}$}  & & \multicolumn{1}{c}{$\Delta x$} & \multicolumn{1}{c}{$\Delta y$} &  & \multicolumn{1}{c}{$V_{\rm x}$} & \multicolumn{1}{c}{$V_{\rm y}$} & \multicolumn{1}{c}{$\Delta|V|/|V| $} \\
        &  (km s$^{-1}$) & \multicolumn{1}{c}{(Jy beam$^{-1}$)} &  & \multicolumn{1}{c}{(mas)} & \multicolumn{1}{c}{(mas)} &  & \multicolumn{1}{c}{(km s$^{-1}$)} & \multicolumn{1}{c}{(km s$^{-1}$)} &                  \\
\hline
& & & & & & & & &\\
\endhead
& & & & & & & & &\\
\hline
\endfoot
& & & & & & & & &\\
\hline
\endlastfoot
1   & 72.58 &202.89 (4)& & $  -64.38 \pm 0.10 $ & $  -60.45 \pm 0.11 $& &  $ -10.8 \pm 3.5$ & $  3.0 \pm 3.6$ & 31\% \\
2   & 78.48 &22.21  (3)& & $  437.05 \pm 0.11 $ & $   74.67 \pm 0.28 $& &  $  11.6 \pm 3.5$ & $-18.1 \pm6.3 $ & 26\% \\
3   & 79.32 &19.68  (3)& & $  437.16 \pm 0.10 $ & $   73.42 \pm 0.14 $& &  $  10.0 \pm 3.5$ & $ -0.0 \pm 5.0$ & 36\% \\
4   & 75.95 &15.70  (1)& & $  425.22 \pm 0.10 $ & $  159.39 \pm 0.12 $& &  $   9.4 \pm 3.4$ & $ -0.1 \pm 4.3$ & 37\% \\
5   & 73.21 &12.70  (1)& & $  -61.49 \pm 0.26 $ & $  -76.84 \pm 0.14 $& &  $ -22.3 \pm11.7$ & $-42.8 \pm 8.1$ & 19\% \\
6   & 79.32 & 9.44  (3)& & $  -60.71 \pm 0.08 $ & $  -77.23 \pm 0.09 $& &  $ -51.0 \pm 4.6$ & $-30.3 \pm 5.1$ & 8\% \\
7   & 78.26 & 9.21  (1)& & $  425.12 \pm 0.10 $ & $  158.45 \pm 0.13 $& &  $  10.5 \pm 3.6$ & $-26.0 \pm 4.5$ & 16\% \\
8   & 69.00 & 7.75  (1)& & $  -65.47 \pm 0.10 $ & $  -60.76 \pm 0.12 $& &         ...       &       ...       & ...  \\
9   & 66.47 & 5.14  (4)& & $  -66.17 \pm 0.13 $ & $  -60.82 \pm 0.12 $& &  $   3.4 \pm 4.2$ & $ -5.3 \pm 3.5$ & 60\% \\
10  & 61.83 & 5.08  (4)& & $  -53.18 \pm 0.10 $ & $  -54.25 \pm 0.12 $& &  $ -26.3 \pm 4.0$ & $ 3.5 \pm 4.1 $ & 15\% \\
11  & 77.63 & 5.08  (2)& & $  -59.60 \pm 0.12 $ & $  -77.20 \pm 0.12 $& &  $   1.8 \pm 7.0$ & $-45.5 \pm 8.3$ & 18\% \\
12  & 72.57 & 4.92  (1)& & $  -65.17 \pm 0.10 $ & $  -60.67 \pm 0.14 $& &         ...       &       ...       & ...  \\
13  & 76.58 & 4.82  (4)& & $  425.18 \pm 0.10 $ & $  158.79 \pm 0.11 $& &  $   5.9 \pm 3.3$ & $ -3.6 \pm 4.4$ & 53\% \\
14  & 66.05 & 4.76  (3)& & $  -53.84 \pm 0.10 $ & $  -54.65 \pm 0.11 $& &  $ -22.3 \pm 3.5$ & $ 4.1 \pm 4.2 $ & 16\% \\
15  & 76.37 & 4.17  (3)& & $  -53.29 \pm 0.09 $ & $  -30.59 \pm 0.13 $& &         ...       &       ...       & ...  \\
16  & 72.58 & 3.86  (1)& & $  -62.93 \pm 0.09 $ & $  -60.06 \pm 0.11 $& &         ...       &       ...       & ...  \\
17  & 77.21 & 3.06  (4)& & $  -61.51 \pm 0.08 $ & $  -78.62 \pm 0.09 $& &         ...       &       ...       & ...  \\
18  & 74.26 & 2.80  (2)& & $  -59.39 \pm 0.11 $ & $  -77.00 \pm 0.14 $& &  $  -4.2 \pm 7.0$ & $-46.0 \pm 9.2$ & 20\% \\
19  & 78.90 & 2.54  (4)& & $  -55.73 \pm 0.11 $ & $ -102.53 \pm 0.13 $& &         ...       &       ...       & ...  \\
20  & 75.10 & 2.37  (3)& & $   32.66 \pm 0.08 $ & $  -14.48 \pm 0.09 $& &         ...       &       ...       & ...  \\
21  & 79.74 & 2.27  (3)& & $  420.54 \pm 0.10 $ & $  126.25 \pm 0.12 $& &  $   5.4 \pm 3.6$ & $ -2.5 \pm 3.7$ & 62\% \\
22  & 56.99 & 2.18  (1)& & $  -52.03 \pm 0.10 $ & $  -53.90 \pm 0.13 $& &         ...       &       ...       & ...  \\
23  & 78.69 & 2.10  (1)& & $  -52.17 \pm 0.15 $ & $  -31.90 \pm 0.18 $& &         ...       &       ...       & ...  \\
24  & 71.52 & 1.95  (2)& & $          0       $ & $         0        $& &  $       0      $ & $      0      $ & ...  \\
25  & 78.26 & 1.72  (4)& & $  435.51 \pm 0.10 $ & $   77.83 \pm 0.13 $& &  $   6.0 \pm 3.6$ & $ -3.4 \pm 4.0$ & 54\% \\
26  & 65.63 & 1.64  (1)& & $  -66.53 \pm 0.14 $ & $  -60.98 \pm 0.11 $& &  $ -29.5 \pm 4.7$ & $12.9 \pm 4.1 $ & 14\% \\
27  & 68.36 & 1.52  (1)& & $  -64.97 \pm 0.13 $ & $  -60.60 \pm 0.12 $& &         ...       &       ...       & ...  \\
28  & 81.21 & 1.43  (3)& & $  -62.98 \pm 0.08 $ & $  -77.14 \pm 0.09 $& &         ...       &       ...       & ...  \\
29  & 79.95 & 1.41  (4)& & $  -63.87 \pm 0.13 $ & $ -163.86 \pm 0.10 $& &  $ -18.6 \pm 7.3$ & $-35.6 \pm 5.6$ & 15\% \\
30  & 78.05 & 1.29  (1)& & $  -49.12 \pm 0.14 $ & $ -160.46 \pm 0.11 $& &  $  -6.5 \pm 9.1$ & $-14.4 \pm10.9$ & 67\% \\
31  & 72.37 & 1.09  (1)& & $  -62.09 \pm 0.09 $ & $  -76.67 \pm 0.11 $& &         ...       &       ...       & ...  \\
32  & 80.37 & 0.76  (4)& & $  -57.43 \pm 0.09 $ & $ -162.26 \pm 0.11 $& &         ...       &       ...       & ...  \\
33  & 79.53 & 0.75  (3)& & $  -32.44 \pm 0.10 $ & $  -29.99 \pm 0.12 $& &  $  -4.5 \pm 3.4$ & $-12.4 \pm 3.7$ & 28\% \\
34  & 80.79 & 0.71  (1)& & $  425.21 \pm 0.10 $ & $  158.23 \pm 0.11 $& &         ...       &       ...       & ...  \\
35  & 72.16 & 0.67  (1)& & $  -60.58 \pm 0.10 $ & $  -77.28 \pm 0.11 $& &         ...       &       ...       & ...  \\
36  & 78.48 & 0.59  (2)& & $  -49.46 \pm 0.08 $ & $ -161.70 \pm 0.13 $& &         ...       &       ...       & ...  \\
37  & 66.89 & 0.56  (1)& & $  -66.88 \pm 0.12 $ & $  -61.05 \pm 0.11 $& &         ...       &       ...       & ...  \\
38  & 80.58 & 0.55  (1)& & $  437.41 \pm 0.10 $ & $   67.80 \pm 0.15 $& &  $  11.4 \pm 7.2$ & $ -6.8 \pm10.8$ & 63\% \\
39  & 75.32 & 0.49  (2)& & $ -548.55 \pm 0.09 $ & $ -314.11 \pm 0.11 $& &         ...       &       ...       & ...  \\
40  & 79.53 & 0.48  (2)& & $  -49.38 \pm 0.09 $ & $ -162.86 \pm 0.17 $& &         ...       &       ...       & ...  \\
41  & 76.79 & 0.44  (2)& & $ -548.30 \pm 0.09 $ & $ -314.13 \pm 0.09 $& &         ...       &       ...       & ...  \\
42  & 79.32 & 0.43  (4)& & $ -293.32 \pm 0.08 $ & $ -100.02 \pm 0.08 $& &         ...       &       ...       & ...  \\
43  & 67.94 & 0.42  (1)& & $  -53.79 \pm 0.09 $ & $  -54.67 \pm 0.11 $& &         ...       &       ...       & ...  \\
44  & 80.58 & 0.39  (1)& & $ -287.44 \pm 0.10 $ & $  -98.35 \pm 0.12 $& &  $ -18.6 \pm 3.5$ & $ -1.5 \pm 3.8$ & 19\% \\
45  & 88.59 & 0.38  (1)& & $  -64.55 \pm 0.13 $ & $ -162.29 \pm 0.12 $& &  $  -7.6 \pm 8.2$ & $-34.9 \pm10.7$ & 30\% \\
46  & 63.94 & 0.38  (4)& & $  -54.44 \pm 0.11 $ & $  -54.30 \pm 0.09 $& &         ...       &       ...       & ...  \\
47  & 80.16 & 0.36  (1)& & $  440.10 \pm 0.10 $ & $   58.11 \pm 0.16 $& &  $  18.7 \pm 7.1$ & $-26.3 \pm10.6$ & 30\% \\
48  & 62.47 & 0.35  (4)& & $  -64.15 \pm 0.09 $ & $  -60.07 \pm 0.09 $& &         ...       &       ...       & ...  \\
49  & 79.32 & 0.33  (3)& & $ -293.50 \pm 0.10 $ & $ -100.37 \pm 0.12 $& &  $  -8.8 \pm 3.5$ & $ -8.5 \pm 4.0$ & 31\% \\
50  & 78.26 & 0.32  (2)& & $  -61.72 \pm 0.08 $ & $  -78.38 \pm 0.17 $& &         ...       &       ...       & ...  \\
51  & 62.25 & 0.31  (3)& & $  -54.39 \pm 0.10 $ & $  -54.57 \pm 0.10 $& &         ...       &       ...       & ...  \\
52  & 69.42 & 0.31  (1)& & $  -40.40 \pm 0.10 $ & $   -7.34 \pm 0.11 $& &         ...       &       ...       & ...  \\
53  & 79.74 & 0.31  (1)& & $  284.39 \pm 0.10 $ & $  -55.26 \pm 0.12 $& &  $   0.2 \pm 4.0$ & $-18.3 \pm 4.9$ & 27\% \\
54  & 77.21 & 0.30  (1)& & $  -58.94 \pm 0.10 $ & $  -77.13 \pm 0.13 $& &         ...       &       ...       & ...  \\
55  & 80.16 & 0.28  (1)& & $  437.51 \pm 0.10 $ & $   67.22 \pm 0.20 $& &         ...       &       ...       & ...  \\
56  & 76.16 & 0.27  (1)& & $ -103.16 \pm 0.10 $ & $ -150.22 \pm 0.12 $& &         ...       &       ...       & ...  \\
57  & 89.43 & 0.26  (1)& & $  -49.16 \pm 0.12 $ & $ -158.89 \pm 0.12 $& &         ...       &       ...       & ...  \\
58  & 71.95 & 0.23  (1)& & $  -67.77 \pm 0.10 $ & $  -61.37 \pm 0.11 $& &         ...       &       ...       & ...  \\
59  & 80.58 & 0.22  (4)& & $  -82.16 \pm 0.08 $ & $ -104.88 \pm 0.09 $& &         ...       &       ...       & ...  \\
60  & 77.42 & 0.21  (2)& & $ -546.43 \pm 0.08 $ & $ -314.43 \pm 0.10 $& &         ...       &       ...       & ...  \\
61  & 70.68 & 0.21  (2)& & $  -65.69 \pm 0.08 $ & $  -60.78 \pm 0.09 $& &         ...       &       ...       & ...  \\
62  & 77.21 & 0.20  (1)& & $  418.75 \pm 0.10 $ & $   92.99 \pm 0.12 $& &  $   5.8 \pm 3.8$ & $ -1.6 \pm 4.7$ & 64\% \\
63  & 70.05 & 0.20  (3)& & $   27.36 \pm 0.09 $ & $  -15.57 \pm 0.12 $& &         ...       &       ...       & ...  \\
64  & 59.51 & 0.19  (3)& & $  -57.57 \pm 0.09 $ & $  -63.38 \pm 0.11 $& &         ...       &       ...       & ...  \\
65  & 75.74 & 0.17  (3)& & $ -549.28 \pm 0.09 $ & $ -314.97 \pm 0.11 $& &         ...       &       ...       & ...  \\
66  & 81.63 & 0.16  (1)& & $  -70.96 \pm 0.10 $ & $  -88.63 \pm 0.13 $& &         ...       &       ...       & ...  \\
67  & 79.74 & 0.16  (3)& & $  -53.11 \pm 0.09 $ & $ -105.70 \pm 0.10 $& &         ...       &       ...       & ...  \\
68  & 70.89 & 0.16  (4)& & $  -51.85 \pm 0.09 $ & $  -73.26 \pm 0.10 $& &         ...       &       ...       & ...  \\
69  & 74.68 & 0.15  (4)& & $   31.19 \pm 0.09 $ & $  -13.67 \pm 0.09 $& &         ...       &       ...       & ...  \\
70  & 70.89 & 0.14  (4)& & $  -38.58 \pm 0.09 $ & $  -53.28 \pm 0.09 $& &         ...       &       ...       & ...  \\
71  & 68.57 & 0.14  (4)& & $  -53.90 \pm 0.09 $ & $  -54.56 \pm 0.09 $& &         ...       &       ...       & ...  \\
72  & 67.31 & 0.13  (1)& & $  -63.71 \pm 0.10 $ & $  -60.09 \pm 0.14 $& &         ...       &       ...       & ...  \\
73  & 75.10 & 0.13  (1)& & $ 442.20  \pm 0.10 $ & $  98.34  \pm 0.14 $& &         ...       &       ...       & ...  \\
74  & 75.74 & 0.12  (1)& & $  -51.19 \pm 0.10 $ & $  -68.44 \pm 0.16 $& &         ...       &       ...       & ...  \\
75  & 66.05 & 0.11  (3)& & $ -541.56 \pm 0.10 $ & $ -291.69 \pm 0.12 $& &         ...       &       ...       & ...  \\
76  & 79.53 & 0.11  (4)& & $  -51.08 \pm 0.09 $ & $ -160.20 \pm 0.10 $& &         ...       &       ...       & ...  \\
77  & 73.21 & 0.10  (1)& & $  -67.84 \pm 0.10 $ & $  -61.58 \pm 0.15 $& &         ...       &       ...       & ...  \\
78  & 81.00 & 0.09  (4)& & $  412.27 \pm 0.09 $ & $  171.84 \pm 0.10 $& &         ...       &       ...       & ...  \\
79  & 73.84 & 0.09  (2)& & $  -68.81 \pm 0.09 $ & $  -61.71 \pm 0.15 $& &         ...       &       ...       & ...  \\
80  & 76.16 & 0.09  (3)& & $ -540.64 \pm 0.09 $ & $ -267.93 \pm 0.14 $& &         ...       &       ...       & ...  \\
81  & 77.21 & 0.09  (1)& & $  419.08 \pm 0.10 $ & $  168.99 \pm 0.13 $& &         ...       &       ...       & ...  \\
82  & 71.94 & 0.09  (3)& & $  -99.30 \pm 0.09 $ & $ -173.21 \pm 0.11 $& &         ...       &       ...       & ...  \\
83  & 69.42 & 0.09  (1)& & $  -68.13 \pm 0.11 $ & $  -61.87 \pm 0.15 $& &  $ -26.8 \pm 8.2$ & $ 8.0 \pm 10.6$ & 30\% \\
84  & 76.58 & 0.08  (1)& & $  -52.05 \pm 0.10 $ & $  -68.04 \pm 0.16 $& &         ...       &       ...       & ...  \\
85  & 63.73 & 0.08  (2)& & $  -53.41 \pm 0.09 $ & $  -54.35 \pm 0.14 $& &         ...       &       ...       & ...  \\
86  & 85.85 & 0.07  (4)& & $   -4.96 \pm 0.09 $ & $ -151.45 \pm 0.12 $& &         ...       &       ...       & ...  \\

& & & & & & & & &\\
& & & & & & & & &\\
\multicolumn{10}{c}{Reference Feature: Absolute position \& Proper motion}\\
\hline\hline
Feature & \multicolumn{3}{c}{R.A.(J2000)} & \multicolumn{3}{c}{ Dec.(J2000) } & \multicolumn{1}{c}{$V_{\rm x}$} & \multicolumn{1}{c}{$V_{\rm y}$} & \multicolumn{1}{c}{$\Delta|V|/|V| $} \\
 & \multicolumn{3}{c}{(h m s)} & \multicolumn{3}{c}{(\degr $'$ $''$)} & \multicolumn{1}{c}{(km s$^{-1}$)} & \multicolumn{1}{c}{(km s$^{-1}$)}& \\
\hline
& & & & & & & & &\\
24 &\multicolumn{3}{c}{18:34:40.28950$\pm$0.00010}&\multicolumn{3}{c}{-09:00:38.2044$\pm$0.0015}&$ 5.3 \pm 8.6 $&$ 21.3 \pm 8.6 $& 39\% \\

\end{longtable}
}

\onltab{3}{

\begin{table*}
\centering
\caption{\emph{Parameters of VLBA 1.665~GHz hydroxyl maser features.}
For each identified feature, the label (given in Col.~1) increases with decreasing brightness.
Cols.~2(3)~and~4(5) report the LSR velocity and brightness of the brightest spot for the right (left) circular polarization.
Cols.~6~and~7 give the position offset relative to the feature~\#~1, toward the East and the North directions, respectively.
Cols.~8~and~9 report the fit split of LSR velocity (RCP-LCP) between co-spatial circular polarization components and the inferred l.o.s. magnetic field strength, respectively. The absolute position of the reference feature~\#~1 is reported at the bottom of the Table.} \label{hyd_tab}
\renewcommand{\footnoterule}{}
\begin{tabular}{c c c c c c r r c c c}
\hline \hline
Feature & \multicolumn{2}{c}{$V_{\rm LSR}$ } & \multicolumn{2}{c}{F$_{\rm peak}$}  & & \multicolumn{1}{c}{$\Delta x$} & \multicolumn{1}{c}{$\Delta y$} & & $\Delta V_{\rm z}$$^{a}$ & B$_{\rm z}$  \\
        & \multicolumn{2}{c}{(km s$^{-1}$)} & \multicolumn{2}{c}{(Jy beam$^{-1}$)} & & \multicolumn{1}{c}{(mas)} & \multicolumn{1}{c}{(mas)} & & (km s$^{-1}$)    & (mG) \\
        &     RCP          &       LCP       &     RCP          &       LCP   & &            &           & & & \\
\hline
& & & & & & & & & & \\

1   & 70.88 & 71.76 & 2.39 & 0.40 & &$            0         $ & $            0         $ & & $ -0.86 $ & $ -1.5 $ \\
2   & 67.89 & ...   & 2.02 & ...  & &$ -210.26 \pm 1.07 $ & $  178.74 \pm 0.77 $ & & ... & ... \\
3   & 66.84 & 66.84 & 1.36 & 0.84 & &$   14.24 \pm 0.62 $ & $   18.87 \pm 0.77 $ & & ... & ... \\
4   & 73.87 & 73.87 & 1.01 & 0.16 & &$ -251.52 \pm 0.78 $ & $   88.39 \pm 1.47 $ & & ... & ... \\
5   & ...   & 74.75 & ...  & 0.97 & &$ -385.08 \pm 1.46 $ & $  203.14 \pm 1.00 $ & & ... & ... \\
6   & ...   & 70.57 & ...  & 0.89 & &$  199.12 \pm 1.45 $ & $  174.08 \pm 1.13 $ & & ... & ... \\
7   & 68.07 & ...   & 0.74 & ...  & &$ -185.97 \pm 0.67 $ & $  191.75 \pm 1.29 $ & & ... & ... \\
8   & 72.99 & 72.99 & 0.67 & 0.26 & &$   57.29 \pm 0.71 $ & $  119.21 \pm 0.94 $ & & ... & ... \\
9   & ...   & 75.10 & ...  & 0.60 & &$ -297.88 \pm 1.46 $ & $  130.40 \pm 1.14 $ & & ... & ... \\
10  & 74.92 & ...   & 0.55 & ...  & &$ -358.64 \pm 1.31 $ & $ -170.67 \pm 1.14 $ & & ... & ... \\
11  & 70.35 & 69.82 & 0.52 & 0.23 & &$   56.12 \pm 0.94 $ & $  114.09 \pm 1.10 $ & & $ +0.62  $ & $ +1.0 $ \\
12  & 67.19 & ...   & 0.46 & ...  & &$   45.90 \pm 0.66 $ & $  111.93 \pm 0.98 $ & & ... & ... \\
13  & 76.33 & 79.67 & 0.25 & 0.43 & &$ -223.40 \pm 1.45 $ & $   71.08 \pm 1.20 $ & & $ -3.41  $ & $ -5.8 $ \\
14  & ...   & 76.33 & ...  & 0.41 & &$  366.46 \pm 1.57 $ & $ -149.68 \pm 1.13 $ & & ... & ... \\
15  & 68.77 & ...   & 0.38 & ...  & &$   52.32 \pm 2.18 $ & $  113.99 \pm 1.27 $ & & ... & ... \\
16  & ...   & 63.50 & ...  & 0.31 & &$   32.32 \pm 1.56 $ & $  108.87 \pm 1.73 $ & & ... & ... \\
17  & 64.90 & ...   & 0.30 & ...  & &$   39.23 \pm 1.10 $ & $  111.08 \pm 1.12 $ & & ... & ... \\
18  & 70.18 & ...   & 0.25 & ...  & &$  192.37 \pm 1.33 $ & $  175.05 \pm 1.53 $ & & ... & ... \\
19  & 79.32 & 77.21 & 0.25 & 0.17 & &$ -321.93 \pm 0.90 $ & $  130.00 \pm 2.23 $ & & $ +2.11  $ & $ +3.6 $ \\
20  & 64.55 & 65.08 & 0.23 & 0.13 & &$   37.16 \pm 0.92 $ & $  108.63 \pm 1.33 $ & & $ -0.49  $ & $ -0.8 $ \\
21  & 66.84 & ...   & 0.22 & ...  & &$ -134.66 \pm 0.91 $ & $  147.73 \pm 1.29 $ & & ... & ... \\
22  & 62.62 & 62.79 & 0.16 & 0.20 & &$   26.50 \pm 1.98 $ & $  106.74 \pm 1.81 $ & & ... & ... \\
23  & 69.65 & 69.30 & 0.18 & 0.19 & &$   17.17 \pm 1.78 $ & $   18.23 \pm 2.17 $ & & ... & ... \\
24  & 60.68 & 60.51 & 0.12 & 0.14 & &$   20.64 \pm 1.70 $ & $   70.44 \pm 2.47 $ & & ... & ... \\

& & & & & & & & & & \\
& & & & & & & & & & \\
\multicolumn{11}{c}{Reference Feature: Absolute position}\\
\hline\hline
Feature & \multicolumn{4}{c}{R.A.(J2000)} & \multicolumn{4}{c}{ Dec.(J2000) } & &  \\
        & \multicolumn{4}{c}{(h m s)}     & \multicolumn{4}{c}{(\degr $'$ $''$)}  & &  \\
\hline
& & & & & & & & & & \\
1  &\multicolumn{4}{c}{18:34:40.28515$\pm$0.00026}&\multicolumn{4}{c}{-09:00:38.3783$\pm$0.0044} & &  \\
& & & & & & & & & & \\
\hline
& & & & & & & & & & \\

\end{tabular}

\footnotesize{(a) \
A magnetic field of 0.34~mG shifts the emission velocity of 1665~MHz OH masers by 0.2~km~s$^{-1}$}

\end{table*}
}

\onllongtab{4}{
\begin{longtable}{ccrrcrrcrrr}

\caption{\emph{Parameters of EVN 6.7~GHz methanol maser features.}
For each identified feature, the label (given in Col.~1) increases with decreasing brightness.
Cols.~2,~3~and~4 report the LSR velocity and brightness of the brightest spot, and its percent brightness
variability.
Cols.~5~and~6 give the position offset relative to the feature~\#~10, toward the East and the North directions, respectively.
Cols.~7~and~8 report the projected components of the feature proper motion relative to the center of motion (as defined in Sect.~\ref{CH3OH_results}, and identified with label~\#~0), along the East and North directions, respectively.
Col.~9 shows the relative error of the proper motion amplitude. ``S'' is the barycenter of the methanol maser distribution as defined in Sect.~\ref{CH3OH_results}. The absolute position of the reference feature~\#~10 is reported at the bottom of the Table.} \label{met_tab} \\
\hline
\hline
Feature &  $V_{\rm LSR}$ & \multicolumn{1}{c}{F$_{\rm peak}$} & Var.$^a$ & & \multicolumn{1}{c}{$\Delta x$} & \multicolumn{1}{c}{$\Delta y$} &  & \multicolumn{1}{c}{$V_{\rm x}$} & \multicolumn{1}{c}{$V_{\rm y}$} & \multicolumn{1}{c}{$\Delta|V|/|V| $} \\
        &  (km s$^{-1}$) & \multicolumn{1}{c}{(Jy beam$^{-1}$)} & &  & \multicolumn{1}{c}{(mas)} & \multicolumn{1}{c}{(mas)} &  & \multicolumn{1}{c}{(km s$^{-1}$)} & \multicolumn{1}{c}{(km s$^{-1}$)} &                  \\
\hline
 & & & & & & & & & &\\
\endfirsthead
\caption{continued.}\\
\hline \hline
Feature &  $V_{\rm LSR}$ & \multicolumn{1}{c}{F$_{\rm peak}$} & Var.$^a$ & & \multicolumn{1}{c}{$\Delta x$} & \multicolumn{1}{c}{$\Delta y$} &  & \multicolumn{1}{c}{$V_{\rm x}$} & \multicolumn{1}{c}{$V_{\rm y}$} & \multicolumn{1}{c}{$\Delta|V|/|V| $} \\
        &  (km s$^{-1}$) & \multicolumn{1}{c}{(Jy beam$^{-1}$)} & &  & \multicolumn{1}{c}{(mas)} & \multicolumn{1}{c}{(mas)} &  & \multicolumn{1}{c}{(km s$^{-1}$)} & \multicolumn{1}{c}{(km s$^{-1}$)} &                  \\
\hline
 & & & & & & & & & &\\
\endhead
 & & & & & & & & & &\\
\hline
\endfoot
 & & & & & & & & & &\\
\hline
\endlastfoot
1  & 74.72 & 146.03 & 52\%  & &$ 144.50 \pm  0.08 $ & $  310.91 \pm  0.09 $& & $ 1.9 \pm 0.9  $ & $ 4.8 \pm 0.9 $ & 19\% \\

2  & 75.51 & 113.88 & 21\%  & & $177.17 \pm  0.07 $ & $  233.91 \pm  0.08 $& &        ...       &       ...       & ... \\

3  & 74.28 & 103.74 & 76\% & &$ 346.87 \pm  0.08 $ & $  425.90 \pm  0.08 $& & $ 3.6 \pm 0.9 $  & $  -1.2 \pm 0.9 $ & 24\% \\

4  & 74.98 & 102.28 & 12\%  & &$ 186.40 \pm  0.08 $ & $  257.95 \pm  0.08 $& & $ 0.7 \pm 0.8 $  & $ 7.7 \pm 0.9 $ & 12\% \\

5  & 74.81 &  80.64 & 58\%  & &$ 168.17 \pm  0.08 $ & $  241.89 \pm  0.09 $& &        ...       &       ...       & ... \\

6  & 80.69 & 53.26 & 62\%   & &$ -47.98 \pm  0.08 $ & $   43.87 \pm  0.08 $& & $ -2.4 \pm 0.8 $ & $ 8.2 \pm 0.9 $ & 11\% \\

7  & 75.86 & 47.43 & 24\%   & &$ 176.27 \pm  0.08 $ & $  239.05 \pm  0.08 $& &        ...       &       ...       & ... \\

8  & 81.74 & 37.47 & 18\%   & &$   1.20 \pm  0.07 $ & $    1.27 \pm  0.08 $& & $ -3.4 \pm 0.8 $ & $ -0.9 \pm 0.8 $ & 23\% \\

9  & 74.02 & 36.93 & 93\%  & &$ 156.85 \pm  0.08 $ & $  304.35 \pm  0.09 $& & $  4.2 \pm 0.9 $ & $ 4.3 \pm 1.0 $ & 16\% \\

10 & 82.36 & 36.19 & 6\%    & & $         0       $ & $        0          $& & $ -5.7 \pm 0.8 $ & $ -0.2 \pm 0.8 $ & 14\% \\

11 & 72.70 & 35.76 & 21\%   & &$ 273.15 \pm  0.08 $ & $  277.34 \pm  0.08 $& & $  3.2 \pm 0.8 $ & $  -3.8 \pm 0.9 $ & 18\% \\

12 & 81.21 & 34.26 & 26\%   & &$ -37.21 \pm  0.07 $ & $   61.77 \pm  0.08 $& & $ -3.6 \pm 0.8 $ & $ -3.1 \pm 0.9 $ & 18\% \\

13 & 74.45 & 25.86 & 56\%   & &$ 213.79 \pm  0.09 $ & $  278.25 \pm  0.11 $& &        ...       &       ...       & ... \\

14 & 74.28 & 22.88 & 40\%   & &$ 220.58 \pm  0.08 $ & $  275.38 \pm  0.10 $& &        ...       &       ...       & ... \\

15 & 74.63 & 21.54 & 70\%  & &$ 309.42 \pm  0.12 $ & $  425.16 \pm  0.16 $& & $ 15.9 \pm 1.6 $ & $  -1.6 \pm 1.9 $ & 10\% \\

16 & 79.46 & 20.61 & 46\%   & &$ -34.98 \pm  0.08 $ & $   47.44 \pm  0.08 $& & $ -3.2 \pm 0.9 $ & $ 5.4 \pm 0.9 $ & 15\% \\

17 & 73.84 & 15.77 & 7\%    & &$ 181.98 \pm  0.08 $ & $  272.74 \pm  0.09 $& & $  0.7 \pm 0.9 $ & $ -1.5 \pm 1.0 $ & 60\% \\

18 & 73.75 & 13.58 & 60\%   & &$ 244.61 \pm  0.09 $ & $  276.21 \pm  0.10 $& & $ 12.1 \pm 1.0 $ & $ 3.2 \pm  1.2 $ & 9\% \\

19 & 73.49 & 13.43 & 58\%   & &$ 167.14 \pm  0.08 $ & $  281.57 \pm  0.09 $& & $ -2.6 \pm 0.9 $ & $ -7.1 \pm 1.0 $ & 13\% \\

20 & 74.19 & 12.90 & 60\%   & &$ 194.21 \pm  0.09 $ & $  272.49 \pm  0.13 $& &        ...       &       ...       & ... \\

21 & 76.74 & 12.77 & 19\%   & &$ 190.23 \pm  0.07 $ & $  229.99 \pm  0.08 $& & $ 10.0 \pm 0.8 $ & $ -1.2 \pm 0.9 $  & 8\% \\

22 & 74.89 & 12.28 & 73\%  & &$ 344.95 \pm  0.19 $ & $  418.82 \pm  0.21 $& & $ 14.6 \pm 2.7 $ & $  -5.9 \pm 2.7 $ & 17\% \\

23 & 74.37 & 11.81 & 86\%   & &$ 205.90 \pm  0.12 $ & $  267.36 \pm  0.19 $& &        ...       &       ...       & ... \\

24 & 74.02 & 10.06 & 9\%    & &$ 236.52 \pm  0.09 $ & $  289.86 \pm  0.11 $& &        ...       &       ...       & ... \\

25 & 80.60 & 10.01 & 36\%   & &$ 331.59 \pm  0.08 $ & $ -143.43 \pm  0.09 $& & $  4.3 \pm 1.0 $ & $ -6.8 \pm 1.1 $ & 14\% \\

26 & 73.23 &  9.75 & 41\%   & &$  73.55 \pm  0.08 $ & $  341.05 \pm  0.09 $& & $  -3.1 \pm 0.9 $ & $ 7.0 \pm 1.0 $ & 13\% \\

27 & 79.28 & 8.23 & 39\%    & &$ 245.63 \pm  0.08 $ & $  -54.11 \pm  0.09 $& & $ 3.0 \pm 0.9 $ & $ -5.5 \pm 1.0 $ & 16\% \\

28 & 81.57 & 8.19 & 13\%    & &$ -49.08 \pm  0.08 $ & $  103.30 \pm  0.09 $& & $  -6.9 \pm 0.9 $ & $ 2.4 \pm 1.0 $ & 13\% \\

29 & 73.23 & 6.81 & 78\%    & &$ 220.62 \pm  0.08 $ & $  223.93 \pm  0.10 $& & $ 3.3 \pm 1.0 $ & $  -1.2 \pm 1.1 $ & 29\% \\

30 & 74.72 & 6.57 & ...      & &$ 191.94 \pm  0.21 $ & $  279.13 \pm  0.22 $& &        ...       &       ...       & ... \\

31 & 73.23 & 6.23 & 50\%    & &$ 315.81 \pm  0.08 $ & $  325.45 \pm  0.09 $& & $ 1.3 \pm 1.0 $ & $ -2.0 \pm 1.2 $ & 48\% \\

32 & 74.10 & 5.86 & 35\%    & &$ 228.71 \pm  0.11 $ & $  284.84 \pm  0.16 $& &        ...       &       ...       & ... \\

33 & 80.86 & 5.15 & 101\%   & &$   4.59 \pm  0.11 $ & $    8.87 \pm  0.15 $& &        ...       &       ...       & ... \\

34 & 77.26 & 4.35 & 31\%    & &$ 196.56 \pm  0.08 $ & $  231.42 \pm  0.09 $& &        ...       &       ...       & ... \\

35 & 74.28 & 4.33 & ...      & &$ 191.09 \pm  0.14 $ & $  289.30 \pm  0.25 $& &        ...       &       ...       & ... \\

36 & 74.10 & 4.30 & ...      & &$ 183.95 \pm  0.11 $ & $  258.93 \pm  0.16 $& &        ...       &       ...       & ... \\

37 & 82.62 & 3.23 & 15\%    & &$ -53.78 \pm  0.08 $ & $   40.29 \pm  0.09 $& & $  -4.3 \pm 1.0 $ & $ 5.2 \pm 1.2 $ & 17\% \\

38 & 83.23 & 3.18 & 38\%    & &$ -49.41 \pm  0.08 $ & $   97.93 \pm  0.09 $& & $ -5.5 \pm 0.9 $ & $ 1.3 \pm 1.0 $ & 17\% \\

39 & 80.25 & 2.81 & 39\%    & &$ 313.27 \pm  0.08 $ & $ -135.59 \pm  0.09 $& & $ 0.1 \pm 1.0 $ & $ -6.7 \pm 1.2 $ & 18\% \\

40 & 73.31 & 2.65 & 135\%    & &$ 185.38 \pm  0.09 $ & $  209.75 \pm  0.11 $& &        ...       &       ...       & ... \\

41 & 73.93 & 2.58 & ...      & &$ 176.30 \pm  0.13 $ & $  260.13 \pm  0.29 $& &        ...       &       ...       & ... \\

42 & 72.79 & 2.27 & 29\%    & &$ 262.25 \pm  0.10 $ & $  221.26 \pm  0.15 $& &        ...       &       ...       & ... \\

43 & 81.48 & 2.06 & 77\%    & &$ -51.58 \pm  0.09 $ & $   39.14 \pm  0.12 $& &        ...       &       ...       & ... \\

44 & 78.93 & 1.94 & 24\%    & &$  78.24 \pm  0.08 $ & $  -52.01 \pm  0.09 $& & $  -4.2 \pm 1.0 $ & $ 1.4 \pm 1.2 $ & 24\% \\

45 & 73.58 & 1.85 & ...      & &$ 252.27 \pm  0.11 $ & $  281.21 \pm  0.19 $& &        ...       &       ...       & ... \\

46 & 81.04 & 1.79 & 11\%    & &$ -45.07 \pm  0.09 $ & $  -31.48 \pm  0.12 $& & $  -8.1 \pm 1.1  $ & $  0.1 \pm 1.5 $ & 15\% \\

47 & 78.84 & 1.78 & 124\%   & &$ -145.32 \pm 0.15 $ & $  294.05 \pm  0.23 $& & $ -5.7 \pm 1.8 $ & $ 11.4 \pm 2.5 $ & 19\% \\

48 & 73.23 & 1.72 & 136\%   & &$ 117.85 \pm  0.26 $ & $  315.86 \pm  0.44 $& &         ...       &       ...       & ... \\

49 & 72.87 & 1.71 & 28\%    & &$ 331.36 \pm  0.12 $ & $  326.56 \pm  0.17 $& & $  2.0 \pm 2.0 $ & $  -5.4 \pm 2.5 $ & 42\% \\

50 & 80.42 & 1.65 & 67\%    & &$ 118.38 \pm  0.09 $ & $  -15.13 \pm  0.12 $& & $ -3.5 \pm 1.7 $ & $ -2.0 \pm 2.1 $ & 45\% \\

51 & 72.17 & 1.29 & 21\%    & &$ 189.08 \pm  0.09 $ & $  215.22 \pm  0.11 $& & $  1.2 \pm 1.2 $ & $ -1.6 \pm 1.4 $ & 65\% \\

52 & 80.95 & 1.18 & ...      & &$ 285.89 \pm  0.16 $ & $  -92.10 \pm  0.17 $& &        ...       &       ...       & ... \\

53 & 78.58 & 1.08 & 76\%   & &$ 361.62 \pm  0.16 $ & $  -60.32 \pm  0.21 $& &        ...       &       ...       & ... \\

54 & 79.90 & 1.04 & 58\%    & &$ 148.46 \pm  0.11 $ & $ -102.60 \pm  0.16 $& & $ -3.3 \pm 1.5 $ & $ -11.1 \pm 2.0 $ & 17\% \\

55 & 70.24 & 0.91 & 14\%    & &$ 192.80 \pm  0.09 $ & $   47.85 \pm  0.11 $& & $  -2.4 \pm 1.3 $ & $ -5.2 \pm 1.6 $ & 28\% \\

56 & 83.06 & 0.88 & 25\%    & &$ 202.89 \pm  0.10 $ & $   -1.49 \pm  0.13 $& & $ 1.5 \pm 1.7 $ & $ -3.7 \pm 2.1 $ & 51\% \\

57 & 73.58 & 0.84 & ...      & &$ 221.83 \pm  0.13 $ & $  263.46 \pm  0.18 $& &        ...       &       ...       & ... \\

58 & 82.44 & 0.83 & 18\%    & &$ 211.57 \pm  0.25 $ & $  -10.52 \pm  0.33 $& &        ...       &       ...       & ... \\

59 & 78.49 & 0.78 & 32\%    & &$ 365.62 \pm  0.10 $ & $  -53.08 \pm  0.16 $& &        ...       &       ...       & ... \\

60 & 69.80 & 0.76 & 45\%    & &$ -54.83 \pm  0.11 $ & $  303.27 \pm  0.15 $& & $  -5.4 \pm 1.5 $ & $ 10.0 \pm 2.0 $ & 17\% \\

61 & 76.56 & 0.69 & ...      & &$  84.27 \pm  0.24 $ & $  381.13 \pm  0.26 $& &        ...       &       ...       & ... \\

62 & 79.81 & 0.60 & 57\%    & &$ 110.44 \pm  0.16 $ & $  -67.92 \pm  0.24 $& &        ...       &       ...       & ... \\

63 & 80.16 & 0.55 & 40\%    & &$ 109.85 \pm  0.18 $ & $  -72.17 \pm  0.26 $& &        ...       &       ...       & ... \\

64 & 80.16 & 0.54 & 28\%    & &$ 369.64 \pm  0.19 $ & $ -142.39 \pm  0.24 $& & $ -2.2 \pm 3.0 $ & $ -10.4 \pm 3.2 $ & 30\% \\

65 & 78.76 & 0.47 & 7\%     & &$  18.53 \pm  0.14 $ & $  298.80 \pm  0.22 $& & $ 1.4 \pm 2.4 $ & $ 7.4 \pm 3.2 $ & 43\% \\

66 & 76.91 & 0.47 & ...      & &$ 165.18 \pm  0.25 $ & $  -23.41 \pm  0.28 $& &        ...       &       ...       & ... \\

67 & 79.72 & 0.46 & 42\%    & &$ 160.89 \pm  0.16 $ & $  -26.57 \pm  0.27 $& &        ...       &       ...       & ... \\

68 & 78.23 & 0.41 & 18\%    & &$  83.31 \pm  0.13 $ & $  -38.89 \pm  0.18 $& & $  -1.9 \pm 2.3 $ & $ -0.8 \pm 2.8 $ & 114\% \\

69 & 80.25 & 0.37 & 62\%    & &$ 354.72  \pm 0.21 $ & $ -140.73 \pm  0.38 $& &        ...       &       ...       & ... \\

70 & 77.44 & 0.36 & 65\%    & &$ -30.76 \pm  0.14 $ & $   69.63 \pm  0.22 $& &        ...       &       ...       & ... \\

71 & 77.70 & 0.29 & 85\%   & &$ -142.36 \pm 0.26 $ & $  302.09 \pm  0.41 $& & $  -9.4 \pm 3.7 $ & $ 9.1 \pm 5.1 $ & 34\% \\

72 & 78.76 & 0.25 & 20\%    & &$  -1.08 \pm  0.26 $ & $  313.38 \pm  0.38 $& &        ...       &       ...       & ... \\

73 & 78.23 & 0.25 & 23\%    & &$ 167.15 \pm  0.19 $ & $  -25.88 \pm  0.33 $& &        ...       &       ...       & ... \\

74 & 79.20 & 0.24 & 46\%    & &$ -65.58 \pm  0.19 $ & $  102.70 \pm  0.32 $& &        ...       &       ...       & ... \\

75 & 77.88 & 0.22 & 45\%    & &$ -69.95 \pm  0.33 $ & $  103.10 \pm  0.44 $& &        ...       &       ...       & ... \\

76 & 78.14 & 0.19 & 40\%    & &$ 200.35 \pm  0.35 $ & $  178.88 \pm  0.46 $& &        ...       &       ...       & ... \\

77 & 78.14 & 0.17 & 54\%    & &$ -68.92 \pm  0.37 $ & $  103.28 \pm  0.62 $& &        ...       &       ...       & ... \\

78 & 77.35 & 0.16 & 47\%    & &$ -151.21 \pm 0.28 $ & $  303.05 \pm  0.43 $& &        ...       &       ...       & ... \\

79 & 72.35 & 0.13 & 28\%    & &$ 178.61 \pm  0.38 $ & $  159.23 \pm  0.56 $& &        ...       &       ...       & ... \\

80 & 78.32 & 0.12 & 5\%     & &$  50.77 \pm  0.32 $ & $  257.35 \pm  0.50 $& &        ...       &       ...       & ... \\

81 & 67.87 & 0.10 & ...      & &$ 178.94 \pm  0.48 $ & $   49.84 \pm  0.55 $& &        ...       &       ...       & ... \\

& & & & & & & & & &\\
0  &        &   &   & & $ 131.30 \pm 0.07 $ & $ 149.69 \pm 0.08 $ & &  $   0  $     &      $   0  $     & \\
& & & & & & & & & &\\
S  &        &   &   & &         155         &         143         & &               &                   & \\
& & & & & & & & & &\\
& & & & & & & & & &\\
\multicolumn{10}{c}{Reference Feature: Absolute position}\\
\hline\hline
Feature & \multicolumn{3}{c}{R.A.(J2000)} & \multicolumn{3}{c}{ Dec.(J2000) } & & & & \\
        & \multicolumn{3}{c}{(h m s)}     & \multicolumn{3}{c}{(\degr $'$ $''$)}  & & & & \\
\hline
& & & & & & & & & &\\
10 &\multicolumn{3}{c}{18:34:40.27658$\pm$0.00014}&\multicolumn{3}{c}{-09:00:38.4086$\pm$0.0021} & & & &\\

\end{longtable}
\footnotesize{(a) \
Variability (\%) = (F$_{\rm MAX}$ $-$ F$_{\rm MIN}$ of the brightest spot) / ((F$_{\rm MAX}$ + F$_{\rm MIN}$)/2.0) \\  }
}

\end{document}